\documentclass[pdflatex,sn-mathphys-num]{sn-jnl}

\usepackage{graphicx}%
\usepackage{multirow}%
\usepackage{amsmath,amssymb,amsfonts}%
\usepackage{amsthm}%
\usepackage{mathrsfs}%
\usepackage[title]{appendix}%
\usepackage{xcolor}%
\usepackage{textcomp}%
\usepackage{manyfoot}%
\usepackage{booktabs}%
\usepackage{algorithm}%
\usepackage{algorithmicx}%
\usepackage{algpseudocode}%
\usepackage{listings}%
\usepackage{bm}%
\usepackage{tensor}%
\usepackage{physics}%
\usepackage{braket}%
\usepackage{url}%
\usepackage{hyperref}%

\begin{document}

\title[Article Title]{From Hamilton-Jacobi to Bohm: Why the Wave Function Isn’t Just Another Action}


\author[1]{\fnm{Arnaud} \sur{Amblard}}\email{arnaud.amblard@neel.cnrs.fr}

\author[1]{\fnm{Aurélien} \sur{Drezet}}\email{aurelien.drezet@neel.cnrs.fr}
\equalcont{These authors contributed equally to this work.}

\affil[1]{\orgdiv{Univ. Grenoble Alpes, CNRS}, \orgname{Institut Néel}, \orgaddress{\street{25 av. des Martyrs}, \city{Grenoble}, \postcode{38042}, \country{France}}}


\abstract{This paper examines the physical meaning of the wave function in Bohmian mechanics (BM), addressing the debate between causal and nomological interpretations. While BM postulates particles with definite trajectories guided by the wave function, the ontological status of the wave function itself remains contested. Critics of the causal interpretation argue that the wave function’s high-dimensionality and lack of back-reaction disqualify it as a physical entity. Proponents of the nomological interpretation, drawing parallels to the classical Hamiltonian, propose that the wave function is a ``law-like" entity. However, this view faces challenges, including reliance on speculative quantum gravity frameworks (e.g., the Wheeler-DeWitt equation) and conceptual ambiguities about the nature of ``nomological entities". By systematically comparing BM to Hamilton-Jacobi theory, this paper highlights disanalogies between the wave function and the classical action function. These differences---particularly the wave function’s dynamical necessity and irreducibility---support a \textit{sui generis} interpretation, where the wave function represents a novel ontological category unique to quantum theory. The paper concludes that the wave function’s role in BM resists classical analogies, demanding a metaphysical framework that accommodates its non-local, high-dimensional, and dynamically irreducible nature.}




\maketitle

\section{Introduction}\label{sec1}

Bohmian mechanics (BM)\footnote{For more general introductions to BM see for instance \cite{bricmont2016making} \cite{durr2009physics} \cite{sep-qm-bohm}.}, also known as de Broglie-Bohm theory or pilot wave theory, is a heterodox quantum theory that describes matter as being composed of particles occupying a well-defined position at any time. In the case of a system composed of $N$ particles, the position of the $k^{th}$ particle at time $t$ is written $\bm{Q}_k(t)$, with $k=1,\dots,N$. The set of positions simultaneously occupied by the $N$ particles at an instant $t$ define the configuration of the system at time $t$: $\mathbf{Q}(t):=\big(\mathbf{Q_1}(t),\dots,\mathbf{Q_N}(t)\big)\in \mathbb{R}^{3N}$. The actual configuration $\bm{Q}(t)$ can be represented by a point in the configuration space of the system $\Gamma:=\bigl\{\mathbf{q}=(\mathbf{q}_1,\dots,\mathbf{q}_N)|\bm{q}\in\mathbb{R}^{3N}\bigl\}$\footnote{Where the $\bm{q}_k$ ($k=1,\dots,N$) denote the spatial degrees of freedom of the $k^{th}$ particle.}. Or more simply, for a $N$-particle system $\Gamma=\mathbb{R}^{3N}$. The motion of $N$ particles in physical space can then be represented by the trajectory followed by the point $\bm{Q}(t)$ in the configuration space $\Gamma$ of this $N$-particle system. To describe the temporal evolution of $\bm{Q}(t)$, we need to define the Bohmian laws of motion, and that's where the wave function $\psi(\bm{q},t)$ of the system comes into play. Indeed, in BM, the wave function of the system generates a velocity field $\bm{v}^\psi(\bm{q},t)$ on the configuration space of this system\footnote{In the remainder of this article, we'll set the same mass for all particles, i.e. $m_k=m$.}
\begin{equation}\label{velocity field}
    \bm{v}^\psi(\bm{q},t):=\frac{\hbar}{m}Im\left(\frac{ \bm{\nabla}\psi}{\psi}(\bm{q},t)\right).
\end{equation}
The possible trajectories of the system, on configuration space $\Gamma$, are the integral curves $\bm{Q}(t)$ everywhere tangent to $\bm{v}^\psi(\bm{q},t)$. In other words, $\bm{Q}(t)$ is solution of the following ordinary differential equation:
\begin{equation}\label{guiding equation}
     \frac{d\bm{Q}(t)}{dt}:= \bm{v}^\psi(\bm{q},t)\bigg|_{\bm{q}=\bm{Q}(t)},
 \end{equation}
called the guiding equation. If it were possible to know the exact initial position of the system $\bm{Q}_0=\bm{Q}(t=0)$, it would select a unique trajectory for the system.
Hence, in non-relativistic BM, a complete description of the system at $t$ requires the knowledge of its configuration and wave function, at that instant: $(\bm{Q}(t),\psi(\bm{q},t))$.\\
We have already specified the law of motion for particles, yet what about the time evolution of the wave function? In BM, there is no collapse of the wave function, the wave function always evolves according to Schr\"odinger equation
\begin{equation}\label{Schrodinger equation}
    i\hbar \frac{\partial}{\partial t}\, \psi(\bm{q},t) = -\sum_{k=1}^N\frac{\hbar^2}{2m}\grad_k^2\psi(\bm{q},t)+V(\bm{q})\psi(\bm{q},t).
\end{equation}
Hence in BM, the wave function guides the particles while itself being a solution to an equation of motion, namely the Schr\"odinger equation. Fundamentally, the Bohmian dynamics is defined by a system of two differential equations
\begin{equation}\label{Bohmian dynamics}
    \left\{
        \begin{aligned}
         &i\hbar\frac{\partial \psi(\mathbf{q},t)}{\partial t}+\sum_{i=1}^N\frac{\hbar^2}{2m}\frac{\partial^2 \psi(\mathbf{q},t)}{\partial q_i^2} - V(\mathbf{q},t)\psi(\mathbf{q},t)=0\\
         &\frac{d\vb{Q}}{dt}(t)=\left.\frac{\hbar}{m}Im\left(\frac{\grad \psi}{\psi}(\vb{q},t)\right)\right\vert_{\vb{q}=\mathbf{Q}(t)}  
         \end{aligned}
     \right. 
\end{equation}
Together, these two equations (\ref{Bohmian dynamics}) constitute the Bohmian laws of motion. Even if we haven't introduced them this way, these two equations are in fact connected to each others, as the guiding equation (\ref{guiding equation}) can be easily guessed from Schr\"odinger equation (\ref{Schrodinger equation}). As a matter of fact, it is a well known result, that starting from the non-relativistic Schr\"odinger's equation we can derive a local probability density conservation equation, known as the continuity equation:
\begin{equation}\label{continuity equation}
    \frac{\partial\rho}{\partial t} + \bm{\nabla} \cdot \bm{j} = 0,
\end{equation}
where \(\rho=|\psi|^2\) denotes the probability density of presence and
\(\bm{j}=\frac{\hbar}{2mi}\big(\psi^*\bm{\nabla}\psi-\psi\bm{\nabla}\psi^*\big)\) the probability current. Moreover, local conservation equations such as (\ref{continuity equation}) appear in other branches of physics: in hydrodynamics equation (\ref{continuity equation}) describes the local conservation of mass, while in electrodynamics it describes the local conservation of electric charge. Importantly, in hydrodynamics or electrodynamics, the continuity equation (\ref{continuity equation}) can be used to define a velocity field based on the following relationship
\begin{equation}\label{Bohmian scheme}
    \bm{v}=\frac{\bm{j}}{\rho}.
\end{equation}
Equation (\ref{velocity field}) can be reformulated to give the relationship (\ref{Bohmian scheme})
\begin{equation}
    \bm{v}^\psi(\bm{q},t)=\frac{\frac{\hbar}{2mi}\big(\psi^*\bm{\nabla}\psi-\psi\bm{\nabla}\psi^*\big)}{\psi^*\psi}(\bm{q},t)=\frac{\bm{j}}{\rho}(\bm{q},t).
\end{equation}
Thus, like other branches of physics, BM simply postulates that particles move at the velocity $\bm{v}=\bm{j}/\rho$. From this point of view, the guiding equation (\ref{velocity field}) naturally follows from Schr\"odinger's equation (\ref{Schrodinger equation}) \cite{sep-qm-bohm}. As Goldstein and Teufel put it: ``BM is the most naively obvious embedding imaginable of Schr\"odinger’s equation into a completely coherent physical theory!" \cite[pp.\ 11]{goldstein_quantum_1999}.\\
Furthermore, for an $N$-particle system the continuity equation can be expressed as $\partial_t\rho+\sum_{k=1}^N \bm{\nabla}_k.\bm{j}_k=0$, where $\bm{j}_k=\frac{\hbar}{2mi}\big(\psi^*\bm{\nabla}_k\psi-\psi\bm{\nabla}_k\psi^*\big)$ denotes the probability current associated with the $k^\text{th}$ particle and $\bm{\nabla}_k:=\big(\frac{\partial .}{\partial x_k},\frac{\partial .}{\partial y_k},\frac{\partial .}{\partial z_k}\big)$. Then the velocity field guiding the $k^{th}$ particle is defined as $\bm{v}_k^\psi:=\bm{j}_k/\rho$, yielding
 \begin{equation}
     \bm{v}_k^\psi(\bm{q},t):=\frac{\hbar}{m}Im\left(\frac{ \bm{\nabla}_k\psi}{\psi}(\bm{q},t)\right).
 \end{equation}
Note that, the velocity field $\bm{v}_k^\psi(\bm{q},t)$ that guides the motion of the $k^\text{th}$ particle is not defined on the physical space $\mathbb{R}^3$, but on $\Gamma=\mathbb{R}^{3N}$, the configuration space of the whole system. Consequently, for $N$ entangled particles, if we wish to calculate the velocity of the $k^{\text{th}}$ particle at time $t$, we have to evaluate $\bm{v}_k^\psi(\bm{q},t)$ on $\mathbf{Q}(t):=\big(\mathbf{Q_1}(t),\dots,\mathbf{Q_N}(t)\big)$, i.e. the configuration of the $N$-particle system at that time. That is to say
  \begin{equation}\label{guiding equation N}
     \frac{d\bm{Q}_k(t)}{dt}:= \bm{v}_k^\psi\big(\bm{Q}_1(t),\dots,\bm{Q}_N(t),t\big).
 \end{equation}
Consequently, the velocity of the $k^{\text{th}}$ particle at a given $t$ not only depends on the position of this particle at that instant $\bm{Q}_k(t)$, but also on the positions occupied by all other entangled particles at that instant\footnote{This isn't true if the wave function of the system $\psi(\bm{q})$ is a product state, i. e. if it has the form $\psi(\bm{q})=\psi_1(\bm{q}_1)\psi_2(\bm{q}_2)\dots\psi_N(\bm{q}_N)$.}, thus highlighting the non-local character of Bohmian dynamics.\\
From this brief introduction to BM, it's clear that a complete description of a quantum system at a given instant $t$ requires the knowledge of the system's configuration at that instant, as well as its wave function $(\bm{Q}(t),\psi(\bm{q},t))$. While the physical meaning of the variable $\bm{Q}(t)$ is clear, the physical meaning of the wave function is more mysterious.\\

The remainder of the paper is organized as follows: In Section \ref{sec2}, we explore the question of the physical meaning of the wave function in BM, highlighting its dual probabilistic and dynamical roles. Section \ref{sec3} introduces the causal interpretation of BM, where the wave function is treated as a real physical field that causally acts on particles, while acknowledging its conceptual challenges, such as high-dimensionality and the absence of back-reaction. Section \ref{sec4} critically examines the nomological interpretation, which reinterprets the wave function as a law-like entity by drawing parallels to the classical Hamiltonian. Section \ref{sec5} evaluates key objections to this view, including its reliance on speculative quantum gravity frameworks and its failure to resolve ontological ambiguities. Section \ref{sec6} systematically compares BM to Hamilton-Jacobi theory, proposing the classical action $S$ as a new analogy for the wave function. Section \ref{sec7} identifies critical disanalogies between these frameworks, particularly the wave function’s dynamical necessity and irreducibility. Building on these insights, Section \ref{sec8} advances an argument for a realist interpretation of the wave function, emphasizing its epistemic indispensability in BM. Finally, Section \ref{sec9} concludes that the wave function resists reduction to classical categories, advocating instead for a \textit{sui generis} interpretation that recognizes its ontological novelty.

\section{The Question of the Physical Meaning of the Wave Function in Bohmian Mechanics}\label{sec2}

The question of the physical meaning of the wave function is not specific to BM. This question naturally arises when one considers the empirical success of the quantum formalism, along with the fact that the wave function $\psi(\bm{q},t)$ is the central mathematical object of orthodox quantum mechanics. Its time evolution is ruled by Schr\"odinger's equation (\ref{Schrodinger equation}), while the Born rule allows us to extract a whole range of probabilistic predictions from the wave function. Given the empirical success of the quantum formalism, it seems legitimate, for any physicist or philosopher of a realist streak, to wonder about the physical meaning of the mathematical object $\psi(\bm{q},t)$. This metaphysical question\footnote{Note that we will use the words metaphysics and ontology as synonyms in this paper.} has been the subject of intense debates in the field of the foundations of quantum mechanics during the last two decades (see \cite{chen2019realism} for a summary of this debate and \cite{ney_wave_2013} for detailed presentations of the different positions).\\
While the question of the physical meaning of the wave function is not specific to BM, this theory offers a specific theoretical framework in which to conduct this reflection. First, in BM, as in orthodox quantum mechanics, the wave function plays a probabilistic role. In quantum equilibrium, the probability of measuring the position of a particle whose wave function is $\psi(\bm{q},t)$ in the infinitesimal volume $d^3x$: $P(\bm{q}\in d^3x,t)=|\psi(\bm{q},t)|^2d^3x$. In other words, in BM, according to the \textit{quantum equilibrium hypothesis} the probability density  $\rho$ for a system to be in configuration $\bm{q}$ is always given by the Born rule $\rho(\bm{q},t)=|\psi(\bm{q},t)|^2$, even outside measurements. On the other hand, in BM the wave function also plays a dynamical role by guiding the particles, as described by the guiding equation (\ref{guiding equation}). Hence from a Bohmian perspective the discussion of physical meaning of the wave function must take into account the dynamical role played by the wave function in BM. In this context, questioning the meaning of the wave function takes a different turn, since it largely consists in questioning the type of physical interaction described by the guiding equation (\ref{guiding equation}).\\
A first possible interpretation of the guiding equation is the causal interpretation. Indeed, as illustrated by the title of the famous Bohmian textbook written by Peter Holland: `The Quantum Theory of Motion, an Account of the de Broglie-Bohm Causal Interpretation of Quantum Mechanics' (\cite{holland1995quantum}), the de Broglie-Bohm theory has long been interpreted in a causal manner. While more sophisticated versions of the causal interpretation have been developed (see, for instance, \cite{valentini1992pilot}), in the next section, we introduce a somewhat naive causal interpretation, which nevertheless likely represents the starting point of many Bohmians' intellectual journey.

\section{The Causal Interpretation of BM}\label{sec3}

 BM is usually introduced in the context of the double-slit experiment. The seemingly paradoxical results of this experiment are simply explained by the existence of a wave $\psi$ that pushes the particle while propagating in physical space. According to this explanation of the double slit experiment, the Bohmian ontology is made of two elements: a physical field described by the wave function $\psi$, and particles described by their position $\bm{Q}(t)$. Such an interpretation of BM can be described as causal \cite[pp.64]{matarese_metaphysics_2023}, insofar as the guiding equation (\ref{guiding equation}) describes the causal action of the wave $\psi$ on the particles. For the wave function to be able to have a causal effect on particles, it must represent a genuine physical entity, that's why the causal interpretation of BM is closely tied to \textit{wave function realism}, i.e., a specific interpretation of the wave function as describing a real physical field. For instance, John Bell wrote about de Broglie-Bohm theory:
\begin{quote}
    Note that in this compound dynamical system the wave is supposed to be just as ‘real’ and ‘objective’ as say the fields in classical Maxwell theory–although its action on the particles is rather original. No one can understand this theory until he is willing to think of $\Psi$ as a real objective field rather than just a `probability amplitude.’ Even though it propagates not in 3-space but in 3N-space. \cite[pp.123]{bell_speakable_1987}
\end{quote}
Despite the ability of the causal interpretation to explain the outcomes of the double-slit experiment in a very simple way, this interpretation comes up against two objections \cite[pp.9]{durr_bohmian_1995}: 
\begin{itemize}
    \item[(1)] As pointed out by Bell in the previous quotation, the $\psi$ wave does not propagate in physical space, but in configuration space, a highly dimensional space. In Cartesian coordinates, the wave function of a system of $N$ particles is written as $\psi(x_1,y_1,z_1,\dots,x_N,y_N,z_N)$ and is therefore defined on the configuration space of the system $\Gamma=\mathbb{R}^{3N}$, and not on the mathematical representation of physical space: $\mathbb{R}^3$. The $\psi$ function associates a specific complex value with each point in the configuration space, and is therefore a high dimensional field. However, contrary to a physical field, the wave function does not take a definite value at any single point of the physical space\footnote{For an $N$-particle system with a wave function $\psi(x_1,y_1,z_1,\dots,x_N,y_N,z_N)$, selecting a specific point of physical space and plugging its Cartesian coordinates $(X,Y,Z)$ in the wave function yields $\psi(x_1=X,y_1=Y,z_1=Z,x_2,y_2,z_2,\dots,x_N,y_N,z_N)$, which as no specific value since the value of $(x_k,y_k,z_k)$ is left unspecified for $k=2,\dots,N$. Hence this wave function has no definite value at any point of the physical space.}. Consequently, the wave function cannot be directly identified with a physical field \footnote{Note that, the wave function is sometimes defined as a \textit{multi-field} on physical space \cite{hubert_romano}. Despite being physical, a \textit{multi-field} is not a usual physical field.}.
    \item[(2)] The causal interpretation of BM violates the action-reaction principle since, according to this interpretation, the guiding equation (\ref{guiding equation}) describes the action of the wave on the particles, while no equation describes the reaction of the particles on the wave. Indeed, as Dürr Goldstein and Zanghì pointed out: ``[The wave function] evolves autonomously via Schr\"odinger’s equation, in which the actual configuration Q does not appear. Indeed, the actual configuration could not appear in Schr\"odinger’s equation because this equation is also in orthodox quantum theory, and in orthodox quantum theory there is no actual position or configuration." \cite[pp.8]{goldstein_reality_2011}. 
\end{itemize}
Although these two objections have sometimes been presented as objections to BM rather than to its causal interpretation, since the work of Dürr Goldstein and Zanghì (DGZ from now on) in \cite{durr_bohmian_1995}\cite{goldstein_reality_2011}, these two objections have instead been reinterpreted as valuable clues about the nature of the wave function \cite{sep-qm-bohm}. These two objections are indeed the starting point of DGZ's interpretation of the wave function, namely the nomological interpretation.

\section{The Nomological Interpretation of the Wave Function}\label{sec4}

Based on the two previous objections to the causal interpretation, DGZ's approach runs roughly as follows. The wave function being defined on a highly dimensional space, it neither describes a real physical field, nor a material entity. Hence, according to Goldstein and Zanghì, the wave function ``is not an element of physical reality" \cite[pp.95]{ney_wave_2013}. Consequently, the guiding equation does not describe the causal action of the wave function on the particles, and it comes as no surprise that the wave function is not subject to the action-reaction principle. On the other hand, the wave function undeniably plays a central role in the formulation of the Bohmian laws of motion so it must represent something real. Here's what DGZ wrote
\begin{quote}    
    We propose that the wave function belongs to an altogether different category of existence than that of substantive physical entities, and that its existence is nomological rather than material. We propose, in other words, that the wave function is a component of physical law rather than of the reality described by the law. \cite[pp.10]{durr_bohmian_1995}
\end{quote}
DGZ point out that, as strange as it might sound, this situation is in fact nothing new in physics. In CM (classical mechanics), the Hamiltonian of a $N$-particle system reads  $H_{class}(\bm{q}_1,\dots,\bm{q}_N,\bm{p}_1,\dots,\bm{p}_N,t):=H_{class}(\xi,t)$, with $\xi\in\mathbb{R}^{6N}$. From a mathematical point of view, the classical Hamiltonian is also a field on a high-dimensional space,  namely the phase space\footnote{For a $N$-particles system, the phase space is $6N$-dimensional.}. Besides being a high dimensional field, according to Hamilton's equations\footnote{ To be more specific, according to Hamilton's equations, the velocity of particle $i$ is given by the relation $\Dot{q}_i=\partial H_{class}/\partial p_i$ and its acceleration (times its mass) by $\Dot{p}_i=-\partial H_{class}/\partial q_i$.}, the classical Hamiltonian, through its gradient, generates the motion of the classical particles. More precisely, DGZ summarize this dynamical similarity by writing
\begin{eqnarray*}
\frac{d}{dt} \bm{Q}(t)\sim \textrm{Der} \log(\Psi(\bm{Q},t))\longleftrightarrow \frac{d}{dt} \bm{\xi}(t)\sim \textrm{Der} H_{class}(\bm{\xi},t)
\end{eqnarray*}
Furthermore, in statistical mechanics, $H_{class}$ generates probabilistic predictions in a similar way than the wave function in quantum mechanics
\begin{equation*}
    \rho_{class}\sim e^{const.H_{class}} \longleftrightarrow \rho_{quant}\sim |e^{const.log\psi}|,
\end{equation*}
$H_{class}$ would thus be the classical analog of the logarithm of the wave function $log(\Psi(\mathbf{q}))$. DGZ \cite{durr_bohmian_1995} summarize this analogy by writing
\begin{equation*}
    log(\Psi(\mathbf{q})) \longleftrightarrow H_{class}
\end{equation*}
How can this analogy shed light on the physical meaning of the wave function? DGZ explain that, in Hamiltonian mechanics, the state of a system is entirely described by its generalized coordinates ($\bm{q}$) and its conjugate momenta ($\bm{p}$) and the Hamiltonian only makes it possible to apply the laws of motion to this system, i.e. to determine the evolution of the system's $q$'s and $p$'s \cite{durr_bohmian_1995}. Hamiltonian mechanics is not about describing the Hamiltonian, it’s about describing the trajectories of the particles constituting the system. Unlike particles, the Hamiltonian is not treated as a material entity:
\begin{quote}
   Everybody knows that the Hamiltonian is just a convenient device in terms of which the equations of motion can be nicely expressed. We're suggesting that you should regard the wave function in exactly the same way. \cite{goldstein_reality_2011} 
\end{quote}
Hence, as the classical Hamiltonian, the wave function is just a `convenient device' to express the laws of motion obeyed by quantum particles: ``the wave function is a component of physical law rather than of the reality described by the law" \cite[pp.10]{durr_bohmian_1995}. To be more specific the analogy $log(\Psi(\mathbf{q})) \longleftrightarrow H_{class}$ justifies the attribution of the same metaphysical status to the classical Hamiltonian and the wave function: both the wave function and the classical Hamiltonian belongs to the category of nomological entities \cite{goldstein_reality_2011} \cite{allori_primitive_nodate}. This notion makes more sense in the context of the primitivist approach\footnote{For a detailed introduction to \textit{primitive ontology} see \cite{allori_primitive_nodate} \cite[pp.58]{ney_wave_2013}}.\\
As explained by Valia Allori \cite{allori2013primitive}, the primitivist approach is a normative approach holding that, in order to explain our experience of the macroscopic world, all fundamental physical theories must postulate a \textit{primitive ontology}. The \textit{primitive ontology} of a fundamental physical theory lists the elementary constituents of matter and represents them with the help of primitive variables. Since the \textit{primitive ontology} consists of microscopic material entities living in the three-dimensional physical space (or 4-dimensional spacetime), the primitivist approach requires primitive variables to be defined on a three dimensional (or 4-dimensional) mathematical space. In addition, a fundamental physical theory must describe the dynamics of its \textit{primitive ontology}. The formulation of such laws generally requires the intervention of additional variables, i.e. non-primitive variables. This non-primitive variables also have an ontological character, but they differ from primitive variables in that they do not represent material entities. When involved in the formulation of a law, these non-primitive variables are referred to as nomological entities \cite{allori2013primitive}.\\
Therefore, according to the nomological interpretation, in BM, the wave function intervenes as a nomological entity in the guiding equation, while being itself a solution of an equation of motion, the Schr\"odinger equation. Hence, adopting the nomological interpretation leads to a curious situation in which a nomological entity is itself subject to a law of nature. This admittedly somewhat curious situation has been made more dramatic than it is by DGZ's tendency to identify the wave function with a law: ``What it suggests to us is that you should think of the wave function as describing a law, not as some sort of concrete physical reality." \cite{goldstein_reality_2011}. In the founding article of the nomological interpretation, DGZ titled ``The Wave Function as a Law''\cite[pp.10]{goldstein_quantum_1999}, thus tending to identify the notion of a nomological entity with that of a law. However, identifying the wave function with a law of nature makes things worse. Indeed, we end up with the very counter-intuitive situation in which a law of nature (the wave function) is subject to another law of nature (the Schr\"odinger equation).\\
Moreover, since the wave function evolves over time and varies from one system to another, identifying the wave function with a law of nature leads to a description of the world in which a law (the wave function) evolves over time and is not the same for all the systems. The wave function would then describe a contingent and particular (i.e. non-universal) law of nature, thus conflicting with the widespread intuition that laws of nature are, by definition, necessary and universal. Lastly, the laws of nature are supposed to be ``sovereign", i.e. they can't be manipulated or controlled, whereas we can prepare, and thus control, the wave function of a system \cite{goldstein_reality_2011}.\\

In order to restore these three characteristics to the wave function, DGZ draw a fundamental distinction between the wave function of the universe $\Psi_{Un.}$ and subsystem wave functions $\psi$. They argue that only the metaphysical status of $\Psi_{Un.}$ really matters, because subsystem wave functions are derived from the universal wave function\footnote{The Bohmian description of the universe is given by $(\Psi_{Un.},Q_{Un.})$. We can derive from it the wave function of subsystem $\psi_t(x)$, by plugging the actual configuration $Y(t)$ of its environment in the universal wave function: $\psi_t(x)=\Psi_{Un.}(x,Y(t))$. $\psi_t(x)$ is called the conditional wave function, and the velocity of the subsystem is proportional to its gradient $\dot{X}(t)=\frac{\hbar}{m}Im\frac{\nabla_x\psi_t(x)}{\psi_t(x)}\Big|_{x=X(t)}$ \cite[pp.83]{durr2020understanding}.}. Subsystem wave functions are not additional fundamental entities, they are consequences of the universal wave function. Hence, if we can understand the nature of the universal wave function, we automatically resolve metaphysical questions about subsystem wave functions. 
\begin{quote}
    With regard to this difficulty, it is important to recognize that there’s only one wave function we should be worrying about, the fundamental one, the wave function $\Psi$ of the universe. In BM, the wave function $\Psi$ of a subsystem of the universe is defined in terms of the universal wave function $\Psi$. Thus, to the extent that we can grasp the nature of the universal wave function, we should understand as well, by direct analysis, the nature of the objects that are defined in terms of it; in particular, we should have no further fundamental question about the nature of the wave function of a subsystem of the universe. So we focus on the former.\cite{goldstein_reality_2011}
\end{quote}
This focus on the universal wave function is not arbitrary, it arises from the non-separability of entangled quantum states. Indeed, due to quantum entanglement, subsystems generally lack independent wave functions, as their states cannot be factorized into product states. Consequently, the universe is uniquely privileged as the sole system with a well-defined wave function: ``from a fundamental point of view, the only genuine Bohmian system in a Bohmian universe—the only system you can be sure is Bohmian—is the universe itself, in its entirety."\cite{goldstein_reality_2011}.\\
DGZ further posit that $\Psi_{Un.}$ the Wheeler-DeWitt equation
\begin{equation}\label{Wheeler-DeWitt}
    \hat{\mathcal{H}}\Psi_{Un.}=0,
\end{equation}
according to which the wave function of the universe $\Psi_{Un.}$ would be static. From the perspective of the nomological interpretation, ``the situation is rather dramatically transformed" \cite{goldstein_reality_2011}. The universal wave function can't be prepared, and the Wheeler-DeWitt equation offers the advantage of providing a universal, immutable wave function that aligns with our intuitions about the nature of physical laws. DGZ summarize this achievement with the remark that it is ``just what the doctor ordered" \cite[pp.11]{goldstein_reality_2011}.

\section{Challenges to the Nomological View: Quantum Gravity and Ontological Vagueness}\label{sec5}

While the assertion that the nomological interpretation is “just what the doctor ordered” sounds provocative, DGZ's ingenuity in developing and defending this interpretation deserves recognition They construct a coherent framework for interpreting the wave function, directly addressing criticisms that BM relies on an unfamiliar, high-dimensional field$\psi$ guiding particles without back-reaction. DGZ reframe these features as evidence for $\psi$ non-physicality: its high-dimensionality and lack of reciprocal particle interaction align with the mathematical role of a law-like entity. By reclassifying $\psi$ as a nomological entity, they intended to purify Bohmian ontology of the “strange” field while leveraging formal parallels between $\psi$ and the classical Hamiltonian $H_{class}$. This analogy exposes a double standard: physicists accept $H_{class}$ guiding particles despite its high dimensionality, yet reject BM for analogous reasons. Based on the pervasiveness of entanglement, they astutely used Wheeler-DeWitt equation to preempt critiques against the nomological interpretation.\\
Yet for all its ingenuity, the interpretation falters.

\subsection{Quantum Gravity}

Basing Bohmian ontology on the Wheeler-DeWitt equation—a speculative quantum gravity proposal—is fraught with difficulties. While theoretically promising \cite{struyve_bohmian_2024,goldstein_quantum_1999}, the equation remains empirically unverified and faces unresolved challenges: mathematical ambiguities (e.g., factor ordering, regularization), the absence of a well-defined Hilbert space for the universal wave function, and no consensus on handling divergences \cite{giulini2003quantum,r2005solving}. The nomological interpretation requires the universal wave function $\Psi_{Un.}$ to be immutable and unique, as DGZ emphasize: ``This fundamental wave function [...] is static, stationary, and, in the view of many physicists, unique." \cite{durr_bohmian_1995}.  However, the equation alone does not guarantee a unique solution, its solutions depend critically on the imposition of specific boundary conditions, such as the Hartle-Hawking ``no-boundary" proposal or Vilenkin's tunneling condition \cite{PhysRevD.58.067301}. Without such boundary conditions, the Wheeler-DeWitt equation admits an infinite number of solutions, leading to ambiguity in the determination of the universal wave function. This lack of uniqueness undermines the wave function's ability to serve as a universal law, as required by the nomological interpretation. This ambiguity, compounded by disagreement over boundary conditions, validates skepticism toward the interpretation’s coherence.\\
To be clear: this critique does not dismiss the Wheeler-DeWitt equation’s theoretical value. Rather, it underscores the precariousness of anchoring BM’s ontology to an unproven quantum gravity framework---a field marked by profound uncertainty.\\
Yet there are deeper conceptual issues plaguing the nomological interpretation.

\subsection{The Wave Function is not a Law}

A brief examination of BM’s formalism shows that the wave function does not constitute the law governing particle motion—this role belongs to the guiding equation (\ref{guiding equation}).\\
While the metaphysical status of physical laws is the subject of intense philosophical debates for centuries, the form taken by a law of nature within the framework of a fundamental physical theory is crystal-clear. A law of nature invariably takes the form of an equation connecting different physical quantities expressed through variables, functions or operators. It follows that a law of nature cannot be represented by a single mathematical function, such as the wave function, in the formalism of a physical theory. In addition, as pointed out by Vera Matarese: ``laws have truth-values, but the wave function can be neither true nor false." \cite[pp.128]{matarese_metaphysics_2023}. Even if the universal wave function satisfied the Wheeler-DeWitt equation \ref{Wheeler-DeWitt}, this would not transmute it into a law, and calling $\Psi_{Un.}$ a law would imply that a law (i.e. the universal wave function) obeys another law (i.e. Wheeler-DeWitt equation). In contrast, the guiding equation (\ref{guiding equation}) unambiguously qualifies as a Bohmian law: its mathematical structure is universal, timeless, and immune to manipulation.\\
This distinction seems elementary, and we presume DGZ recognize it. Why then do they assert the wave function’s status as a law or law-like? We posit that the answer resides in the ontological ambiguity surrounding the concept of nomological entities. With no established understanding of what constitutes a nomological entity, it becomes expedient to conflate the wave function with a law.

\subsection{Ontological Vagueness and the Hamiltonian Analogy}

As outlined in Section \ref{sec4}, the notion of nomological entity receives a formal definition in the PO (Primitive Ontology) approach: ``The formalism of the theory contains [...] nonprimitive variables necessary to mathematically implement how the primitive variables will evolve in time." \cite[pp.60]{allori2013primitive}. Despite this formal definition, the metaphysical status of nomological entities remains deeply ambiguous. In particular, we can debate whether nomological entities are ontological or epistemic, what Craig Callender calls the ``it or bit'' debate \cite{callender_one_2015}. Such a debate is obviously connected to philosophical debates about laws of nature, yet, unlike laws of nature---which, despite philosophical disputes, at least align with intuitive notions of governance---nomological entities lack any pre-theoretical intuition. Regarding this metaphysical question, Valia Allori adds (to the previous quote) the following footnote: ``The metaphysical status of such nonprimitive variables is up for debate [...], but surely they do not represent physical objects" \cite[pp.60]{allori2013primitive}, a stance mirroring Goldstein and Zanghì’s assertion that the wave function ``is not an element of physical reality" \cite[pp.95]{ney_wave_2013}. Such statements suggest an epistemic interpretation of the wave function, but proponents of the nomological view might retort that these entities belong to a non-primitive ontology. This response, however, merely relocates the problem: is non-primitive ontology itself part of physical reality? To untangle this, we analyze three possibilities:
\begin{itemize}
    \item[(1)] \textbf{Epistemic Status:} If nomological entities are epistemic constructs, then the wave function dissolves into a calculation tool\footnote{This is suggested by the following quote “After all, [the guiding equation] is an equation of motion, a law of motion, and the whole point of the wave function here is to provide us with the law, i.e., with the right-hand side of this equation” \cite[pp.9]{goldstein_reality_2011}}, leaving Bohmian ontology with only particles. This raises a critical challenge: how does BM explain interference patterns or wave-like phenomena (e.g., the double-slit experiment)? A proponent might argue that the guiding equation is real while denying reality to the wave function itself. Yet this bifurcation leads to a very peculiar explanatory scheme\footnote{Wave-like behavior are classically explained by the occurrence of a wave.}, where wave-like behaviors would occur without being caused by a physical wave. If neither the law (i.e. the guiding equation) nor the nomological entity (i.e. the wave function) is real---as Humeans could argue---then we simply don't see any explanation for the occurrence of interference patterns in QM (nor for the non-locality).
    \item[(2)] \textbf{Ontic Status:} If nomological entities are physical/ontic, then the Bohmian ontology has the necessary ingredients to explain the so called ``wave-particle duality". However, in this case, when applied to the wave function, the category of nomological entity sounds as a strange intermediate step between between classical fields and laws. If the proponents of the nomological interpretation want to keep the Bohmian ontology as close as possible to the classical ontology, as stated by Allori (see quotation bellow), then they have to find a compelling classical analogue to the wave function.
    \item[(3)] \textbf{Neither ontic nor epistemic:} If nomological entities are neither epistemic nor physical/ontic, the concept becomes incoherent.
\end{itemize}
Options (1) and (3) appear untenable. Regarding (1), even DGZ implicitly reject it. For instance, when explaining the double-slit experiment, Goldstein states: ``While each trajectory passes through only one slit, the wave passes through both" \cite{sep-qm-bohm}, implicitly attributing physical reality to the wave---a point we revisit in Section \ref{sec8}.\\
Thus, the wave function likely represents something real, something physical. Yet labeling it a nomological entity invents a new metaphysical category---precisely why intuitions about nomological entity remain absent. This conceptual innovation is in tension with the goal of the PO approach, which, according to Allori, ``reflects the desire to keep the scientific image closer to the classical way of understanding things, given that it is possible" \cite[pp.62]{allori2013primitive}. Once this goal stated, it then becomes a vital issue for the nomological interpretation to make the category of nomological entity look familiar. That's why, proponents of the nomological interpretation lean heavily on the Hamiltonian analogy, which serves dual purposes: (a) familiarizing the wave function by linking it to classical concepts, and (b) grounding intuitions about nomological entities through concrete examples. For these reasons, the analogy between $log(\psi)$ and $H_{class}$ occupies a central place in the defense of the nomological interpretation.\\
However, the analogy between the wave function and the classical Hamiltonian is arguably not the most appropriate one. Admittedly, the classical Hamiltonian guides particles in a similar way to the wave function, but unlike the latter, it is not a solution of a partial differential equation (PDE):
\begin{quote}
    Perhaps the most serious weakness in the analogy is that, unlike $H_{class}$, $\psi = \psi_t$ is time-dependent, and indeed is a solution of what we regard as the fundamental equation of motion for $\psi$, [Schr\"odinger equation]. \cite[pp.11]{durr_bohmian_1995}
\end{quote}
The fact that the classical Hamiltonian is not solution of a PDE, could be dismissed as a superficial flaw in the analogy, a mere formal weakness. Yet this objection obscures a deeper issue: what truly matters is that the wave function evolves according to a wave equation. This mathematical property directly reflects is indispensable for explaining the wave-like phenomena pervasive in quantum mechanics, such as interference patterns. The fact that the wave function is solution of a wave equation should not be neglected when addressing its physical meaning. Moreover, the structural divergence between classical and BM underscores this point: whereas classical systems are fully described by the phase-space variables $(\bm{q},\bm{p)}$, BM necessitates the pair $(\psi,\bm{Q})$. Here, the wave function$\psi$ is not merely a guiding agent but an irreducible component of the physical state---a stark ontological departure from the classical Hamiltonian’s auxiliary role.\\
DGZ are perfectly aware of these limitations when advising readers not to take this analogy too seriously
\begin{quote}
    Now we do not think that this analogy should be taken too seriously or too literally; it’s not a particularly good analogy—but it’s better than it has any right to be. It does, however, have the virtue that it stimulates a new direction of thought concerning the meaning of the wave function, and that is a great virtue indeed. \cite{durr_bohmian_1995}
\end{quote}
Yet, this analogy remains essential to the defense of the nomological interpretation. Without it, the nomological interpretation would lack even a rudimentary conceptual anchor, leaving the wave function’s status wholly enigmatic. Consider, for instance, a hypothetical scenario where DGZ classify the wave function as nomological without invoking any classical analogue—no reference to the Hamiltonian, no illustrative example of a nomological entity. In such a case, how would we conceptualize the nature of a nomological entity? What ontological features could we ascribe to it? Stripped of the Hamiltonian analogy, the category of ``nomological entity" collapses into pure abstraction, devoid of intuitive content. Indeed, without the Hamiltonian analogy, we would have no intuitions whatsoever about nomological entities, and the wave function wouldn't appear a familiar physical entity.\\

Before drawing lessons from these conceptual criticisms, we shall critically examine another possible classical analogue to the wave function. If we are to find a classical analogue to the wave function in BM, we should obviously examine the case of the classical pilot wave theory, namely Hamilton-Jacobi theory.

\section{From Action to Wave Function: Revisiting Classical Pilot-Wave Dynamics}\label{sec6}

Before returning to interpretative questions regarding the nature of the wave function, we are going to put forward a possible new classical analogue to the wave function. Moreover, we emphasize that the analogy between the wave function and the classical action arises independently of ontological commitment regarding the wave function. As we shall see in this section, this analogy is grounded in the mathematical parallels between BM and Hamilton-Jacobi formalism---a formalism that emerges naturally as BM's classical limit, unlike Hamiltonian mechanics. While Holland \cite{holland1995quantum} pioneered this comparison\footnote{Bohm and Hiley repeatedly applied HJ formalism to BM, see for instance \cite{bohm1952suggested}\cite{bohm2006undivided}.}, contemporary discussions often overlook these structural parallels. To rectify this gap, we systematically re-examine the formal kinship between BM and HJ formalism, bracketing metaphysical debates.

\subsection{Hamilton-Jacobi Formalism}

HJ formalism is a mathematical reformulation of (non-relativistic) CM. It originates from Hamilton's principle: if we consider the set $\mathcal{T}$\footnote{where $\mathcal{T}:=\bigl\{q:t\in[t_0,t_1]\rightarrow q(t)\in \mathbb{R}^{3N}\,\big|\, q(t_0)=q_0, q(t_1)=q_1\bigr\}$ is the path space, i.e. the space of all possible paths between $q_0$ and $q_1$ (with $N$ the number of particles).} of all possible paths that could be taken by a physical system between two fixed points $q_0:=q(t_0)$ and $q_1:=q(t_1)$ (i.e. fixed boundary conditions), Hamilton's principle selects the physical paths. According to Hamilton's principle, the physical paths are the ones which extremize the action functional
\begin{equation}\label{Hamilton's principle}
    \frac{\delta I[q(t)]}{\delta q(t)}=0.
\end{equation}
Where the action functional\footnote{A functional maps functions to real numbers, whereas a function typically maps numbers to real numbers. Here, $I:\mathcal{T}\rightarrow\mathbb{R}$ is a functional because its argument is the entire path $q(t)$ over the interval $[t_0,t_1]$, not merely the value of $q(t)$ at a single instant.} $I:\mathcal{T}\rightarrow\mathbb{R}$ is defined as the temporal integral of the system Lagrangian
\begin{equation}\label{action integral}
    I[q(t)]:=\int_{t_0}^{t_1} L(q,\dot{q},t) dt.
\end{equation}
In order to formulate Hamilton's principle, we consider the motion of a system between two fixed points/configurations $q_0:=q(t_0)$ and $q_1:=q(t_1)$, then calculate the value of the action functional $I[q(t)]$ on each possible path connecting $q_0$ and $q_1$, and finally select the physical paths as being those that extremize $I[q(t)]$. However, one can adopt a different approach than the one taken by Hamilton's principle:
\begin{quote}
    An alternative perspective on the meaning of the Hamilton-Jacobi function $S$ may be gained as follows. [...]  We shall now regard the action function $I$ as a quantity associated with just the actual path traversed by the system and consider the change in $I$ induced by variations in the final coordinate $q$ and time $t$ and the initial coordinate $q_0$, keeping the initial time $t_0$ fixed. \cite[pp.33-34]{holland1995quantum}
\end{quote}
In this alternative approach, rather than considering paths between fixed boundary conditions, we instead vary the boundaries themselves. Then the action $S$ becomes a function of the initial coordinate $q_0$, final coordinate $q$, and time $t$, denoted $S(q_0,q,t)$\footnote{Unlike the first approach, where $S$ is a functional (dependent on the entire path connecting two fixed endpoints), here $S$ is a function of the boundary conditions $S(q_0,q,t)$. This functional-to-function shift reflects the focus on $S$'s dynamical guidance rather than path comparison.}. When studying systems with a fixed (though possibly unknown) initial configuration $q_0$, the notation simplifies to $S(q,t)$, emphasizing the role of $S$ as a field over the system’s current configuration space. For $N$ classical particles, this defines a scalar field $S:\mathbb{R}^{3N}\times\mathbb{R}\rightarrow\mathbb{R}$
\begin{equation}
    S(\mathbf{q},t):=\int_{0}^{t} L(q,\dot{q},t) dt,
\end{equation}
where the integral is evaluated along the actual physical trajectory. Crucially $S(\mathbf{q},t)$ satisfies a partial differential equation
\begin{equation}\label{first HJ}
    \partial_t S(\bm{q},t)+H(\bm{q},\partial_t S,t)=0,
\end{equation}
derived by applying variational methods to infinitesimal path variations under fixed initial conditions. This PDE (\ref{first HJ}) is Hamilton-Jacobi equation, it governs $S$'s spatiotemporal evolution and reflects the vanishing of the transformed Hamiltonian in canonical theory. The LHS of (\ref{first HJ}) refers to the vanishing of the new Hamiltonian in canonical transformation. HJ formalism is often introduced as a mathematical tool for finding the simplest canonical transformation for a given system, namely the canonical transformation for which the new Hamiltonian is equal to zero. However, HJ formalism can be recast as a dynamical framework.\\
As noted by Peter Holland \cite{holland1995quantum}, HJ formalism can be understood as a physical formalism describing particles motion with the help of two differential equations and a field on configuration space, namely the action function $S(\mathbf{q},t)$. As a matter of fact, the gradient of $S$ generates a velocity field $\mathbf{v}(\mathbf{q},t)=\nabla S(\bm{q},t)/m$, for the particles. More specifically, for a $N$-particle system, the velocity of the $i^{\text{th}}$ particle at time $t$ obeys the classical guiding equation
\begin{equation}\label{equation de guidage Hamilton-Jacobi}
    \frac{d \mathbf{Q}_i}{dt}(t)=\left.\frac{1}{m}\bm{\nabla}_i S(\mathbf{q},t)\right\vert_{\mathbf{q}=\mathbf{Q}(t)}.
\end{equation}
Expressed in terms of kinetic energy $K=\sum_{i=1}^N\frac{1}{2m}\big(\bm{\nabla}_i S(\mathbf{q},t)\big)^2$ and potential $V$, the HJ equation (\ref{first HJ}) becomes:
\begin{equation}\label{HJe potential}
\frac{\partial S(\mathbf{q},t)}{\partial t}+\sum_{i=1}^N\frac{1}{2m}\big(\bm{\nabla}_i S(\mathbf{q},t)\big)^2 + V(\mathbf{q},t)=0,
\end{equation}
which looks very similar to Schr\"odinger equation, except that it is a non-linear equation. Thus, in the case of a system made up of $N$ classical particles $\bm{Q}(t)=(\bm{Q}_1(t),\dots,\bm{Q}_N(t))$ of mass $m$, and having an action $S(\bm{q},t)$, HJ formalism enables us to calculate the trajectory of this system by solving the two differential equations
 \begin{equation}\label{HJ theory bis}
    \left\{
        \begin{aligned}
         &\frac{\partial S(\mathbf{q},t)}{\partial t}+\sum_{i=1}^N\frac{1}{2m}\bigg(\frac{\partial S(\mathbf{q},t)}{\partial q_i}\bigg)^2 + V(\mathbf{q},t)=0\\
         &\frac{d \mathbf{Q}_i}{dt}(t)=\left.\frac{1}{m}\frac{\partial S(\mathbf{q},t)}{\partial \mathbf{q}_i}\right\vert_{\bm{q}=\mathbf{Q}(t)}
         \end{aligned}
     \right. 
\end{equation}
In a nutshell, the action function $S(\mathbf{q},t)$ is a high dimensional field on configuration space, whose gradient generates the trajectories of classical particles, and which is the solution of a PDE, namely Hamilton-Jacobi equation. This structure---a configuration-space field guiding particles via its gradient while evolving under a PDE---prompts Anthony Valentini to characterize HJ formalism as a ``classical pilot-wave theory" \cite[pp.11]{valentini1992pilot}. While it may sound a bit exaggerated to call HJ formalism an ``actual physical theory, conceptually and mathematically independent of the usual mechanical formulation" \cite[pp.8]{valentini1992pilot}, it nevertheless provides us with a classical pilot-wave formalism. In HJ formalism classical particles are guided ``by a multidimensional ``guiding field" (or pilot-wave) $S$ which has an autonomous existence in configuration space"\cite[pp.8]{valentini1992pilot}.\\
BM describes analogously governs $N$-particle systems via (\ref{Bohmian dynamics}), substituting $S$ with the wave function $\psi$. The formalization of the systems (\ref{HJ theory bis}) and (\ref{Bohmian dynamics}) already highlights the similarities between the classical action and the wave function, as both formalisms employ a high-dimensional field (either $S$ or $\psi$) that simultaneously evolves under a PDE and guides particles through its spatial derivatives. This structural isomorphism motivates the following analogy
\begin{equation*}
    \psi \longleftrightarrow S_c.
\end{equation*}
This analogy can be further refined by reformulating these two dynamics in one and the same formalism, which might be described as a probabilistic HJ formalism.

\subsection{A Probabilistic Hamilton-Jacobi Formalism}\label{sec6.2}

To reformulate Bohmian dynamics, let us start by writing the wave function (of spinless particles) in polar form $\psi_q(\mathbf{q},t)=R_q(\mathbf{q},t)e^{iS_q(\mathbf{q},t)/\hbar}$. let us note in passing that the function $S_q(\mathbf{q},t)$ has the same dimension as the classical action---energy$\times$time---and therefore describes a quantum action. Firstly, by injecting this expression into (\ref{velocity field}) we re-express the Bohmian guiding equation as $\mathbf{v}_q(\mathbf{q},t)=\frac{1}{m}\frac{\partial S_q(\mathbf{q},t)}{\partial \mathbf{q}}$. Secondly, by injecting the wave function---in polar form---into the Schr\"odinger equation and by splitting the real part from the imaginary part, we perform a Madelung decomposition \cite{madelung1927quantum}. This results in two equations: the quantum equivalent of the Hamilton-Jacobi equation (the first equation of the (\ref{Decomposed Bohmian dynamics}) system) and the continuity equation (the second equation of the (\ref{Decomposed Bohmian dynamics}) system). Bohmian dynamics can then be re-expressed (for $\hbar=1$) as a system of three equations
\begin{equation}\label{Decomposed Bohmian dynamics}
    \left\{
        \begin{aligned}
         &\frac{\partial S_q(\mathbf{q},t)}{\partial t}+\sum_{i=1}^N\frac{1}{2m}\big(\bm{\nabla}_i S_q(\mathbf{q},t)\big)^2 + V(\mathbf{q},t)-\sum_{i=1}^N\frac{\hbar^2}{2m}\frac{\nabla^2R_q(\mathbf{q},t)}{R_q(\mathbf{q},t)}=0\\
         & \frac{\partial R_q(\mathbf{q},t)^2}{\partial t}+\sum_{i=1}^N\frac{1}{m}\bm{\nabla}_i\cdot\big(R_q^2(\mathbf{q},t)\bm{\nabla}_i S_q(\mathbf{q},t)\big)=0 \\
         & \mathbf{v}_{q,i}(\mathbf{q},t)=\frac{1}{m}\bm{\nabla}_i S_q(\mathbf{q},t)
         \end{aligned}
     \right. 
\end{equation}
Strikingly, both BM and HJ formalism employ identical guiding equation:
\begin{equation}\label{classical guiding equation}
    \mathbf{v}_i(\mathbf{q},t)=\frac{1}{m}\bm{\nabla_i}S(\mathbf{q},t).
\end{equation}
In both cases the velocity of the particle is proportional to the action gradient\footnote{Equivalently, particle trajectories are always orthogonal to the action field's level sets.}. This prompt two questions:
\begin{itemize}
    \item[(1)] In the light of this similarity, shouldn't we revise our analogy? Indeed, wouldn't it be more appropriate to establish an analogy between the quantum action and the classical action, rather than between the wave function and the classical action $S_q\longleftrightarrow S_c$?
    \item[(2)] Why do quantum and classical trajectories diverge despite identical guidance equations?
\end{itemize}
To Address these, we recast HJ formalism in the same formalism into a probabilistic framework mirroring the Bohmian dynamics (\ref{Decomposed Bohmian dynamics}). To this end, let us introduce a classical wave function
\begin{equation}\label{Classical WF}
    \psi_c(\mathbf{q},t)=R_c(\mathbf{q},t)e^{iS_c(\mathbf{q},t)/\hbar}.
\end{equation}
Although deterministic, Bohmian dynamics provides a probabilistic description of particle motion. At quantum equilibrium, the probability of a particle's presence is given by the distribution $\rho_q=R_q^2$ and obeys the conservation law described by the second equation of the system (\ref{Decomposed Bohmian dynamics}). By analogy with BM, it is possible to formulate a probabilistic version of HJ theory, in which the amplitude of the classical wave function is related to the probability density of the presence of a classical particle by the relation $\rho_c=:R_c^2$. Since the quantity of particles is conserved over time, the probability density $\rho_c$ must obey the continuity equation $\partial_t\rho_c+\div(\rho_C\bm{v}_c)=0$. We then obtain a probabilistic version of HJ theory that can be described by this three-equation system
\begin{equation}\label{Probabilistic HJ}
    \left\{
        \begin{aligned}
         &\frac{\partial S_c(\mathbf{q},t)}{\partial t}+\sum_{i=1}^N\frac{1}{2m}\big(\bm{\nabla}_i S_c(\mathbf{q},t)\big)^2 + V(\mathbf{q},t)=0\\
         & \frac{\partial R_c(\mathbf{q},t)^2}{\partial t}+\sum_{i=1}^N\frac{1}{m}\bm{\nabla}_i\cdot\big(R_c^2(\mathbf{q},t)\bm{\nabla}_i S_c(\mathbf{q},t)\big)=0 \\
         & \mathbf{v}_{c,i}(\mathbf{q},t)=\frac{1}{m}\bm{\nabla}_i S_c(\mathbf{q},t)
         \end{aligned}
     \right. 
\end{equation}
To deepen our analogy, we can systematically compare the two equation systems obtained\footnote{Interestingly, it is also possible to formulate dynamics (\ref{Probabilistic HJ}) using a nonlinear Schr\"odinger equation describing the time evolution of the classical wave function (see \cite{holland1995quantum} and \cite{callender_one_2015}). This offers another opportunity to formally study the analogies and disanalogies between CM and BM.}. These systems (\ref{Decomposed Bohmian dynamics}) and (\ref{Probabilistic HJ}) are almost identical, differing only in the quantum HJ equation (first equation of (\ref{Decomposed Bohmian dynamics}), which includes an additional term. This additional term corresponds to the quantum potential\footnote{Notably, the appearance of the quantum potential $U^\psi$ in the quantum Hamilton-Jacobi equation is a mathematical fact, which does not compel us to reify it or take a realist stance toward $U^\psi$.} $U^\psi$ and is written as
\begin{equation}\label{Quantum potential}
    U^\psi=-\sum_{i=1}^N\frac{\hbar^2}{2m}\frac{\Delta_iR_q(\mathbf{q},t)}{R_q(\mathbf{q},t)}.
\end{equation}
Given the degree of similarity between the systems (\ref{Decomposed Bohmian dynamics}) and (\ref{Probabilistic HJ}), this difference could, at first sight, seem insignificant. Especially since, in the semi-classical limit, the potential $Q$ cancels out, allowing us to write $S_q \approx S_c$. This might lead us to think that the correct analogy is not between the wave function and the classical action ($\psi \longleftrightarrow S_c$), but between the quantum action and the classical action:
\begin{equation*}
    S_q\longleftrightarrow S_c.
\end{equation*}
However, this would amount to neglecting a crucial difference between quantum action and classical action: while the classical action $S_c$ and the classical amplitude $R_c$ are completely decoupled, this is absolutely not true of the quantum action $S_q$ and the quantum amplitude $R_q$. This difference is explained by the presence of the quantum potential $U^\psi$ in the quantum Hamilton-Jacobi equation (first equation of (\ref{Decomposed Bohmian dynamics})), resulting in the coupling of the quantum action $S_q$ with the quantum amplitude $R_q$. Indeed, the quantum potential $U^\psi$, whose formula is given by (\ref{Quantum potential}) introduces the quantum amplitude into the equation that determines the time evolution of $S_q$, i.e. the quantum Hamilton-Jacobi equation, as a consequence of which the time evolution of $S_q$ depends on $R_q$\footnote{Similarly, $R_q$ depends on $S_q$ through the continuity equation.}. This mutual dependence means $S_q$ and $R_q$ cannot be solved independently, they form a coupled system. The coupling between the modulus and the phase of the wave function has already been presented as the signature of quantum mechanics: ``What is QM? Quantum mechanics is the theory of the interaction between a phase and a modulus” \cite{mauro2003topicskoopmanvonneumanntheory}. As such, $R_q$ is somehow “hidden” behind $S_q$, so that when we write
\begin{equation*}
     S_q\longleftrightarrow S_c,
\end{equation*}
we actually write
\begin{equation*}
    (R_q, S_q)\longleftrightarrow S_c.
\end{equation*}
For spinless particles, we can certainly write the Bohmian velocity field as $\mathbf{v}_{q,i}(\mathbf{q},t)=\frac{1}{m}\frac{\partial S_q(\mathbf{q}, t)}{\partial \mathbf{q}_i}$, i.e. without making $R_q$ appear, but the fact remains that $\bm{v}_q$ implicitly depends on $R_q$. In other words, the amplitude of the quantum wave function $R_q$ plays an implicit role in the Bohmian dynamics\footnote{To be more specific, in BM the $R$ modulus plays a dual role, being both dynamical and epistemic. Indeed, $R$ appears in the conservation law (\ref{continuity equation}) derived from Schr\"odinger's equation, and this law constrains the Bohmian dynamics. Equivariance shows that the equality between the probability density and $R^2$ will be preserved over time, which is fundamental for recovering the statistical predictions of quantum physics using BM. Of course, this so-called quantum equilibrium condition may be relaxed, and the probability density of presence does not necessarily have to be identical to $R^2$ (this is Valentini's research program, which extends that of Bohm and Vigier, also accepted by de Broglie as early as the 1950s). In any case, since the first two equations in (\ref{Decomposed Bohmian dynamics}) are coupled, it would make no sense to attempt to solve the quantum Hamilton-Jacobi equation without taking into account the continuity equation (\ref{continuity equation}).}, which is absolutely not the case for $R_c$ in CM. Because of the coupling of $S_q$ with $R_q$, which is specific to quantum mechanics, we end up with our very first analogy:
\begin{equation*}
    \psi_q \longleftrightarrow S_c.
\end{equation*}
Given all the formal similarities between BM and HJ formalism, the action function appears as better classical analogue for the wave function, than the classical Hamiltonian. We initially thought to have provided a very close analogy between the quantum wave function and the classical action, which would be able to strengthen the nomological interpretation, by classifying the quantum wave function and the classical action in the same category, that of nomological entities. However, contrary to our initial expectations, things are not that simple.

\section{Beyond Analogies: Why the Quantum Wave Function Defies Classical Reduction}\label{sec7}

\subsection{Holland's Example}
As pointed out by Holland \cite{holland1995quantum} and by Callender \cite{callender_one_2015}, a closer look at the respective dynamics of the two theories reveals some differences between the action function and the wave function\footnote{In this section, when we omit the subscript $c$ the letter $S$ refers to the classical action function, which was called $S_c$ in the previous section. On the other hand, the letter $\psi$ only refers to the quantum wave function.}. These differences are illustrated by an example where Holland \cite[pp.36]{holland1995quantum} considers a free classical particle, in which case the Hamilton-Jacobi equation simply reads $\partial_t S+(\nabla S)^2/2m=0$. Solving this equation leads to two different solutions (see appendix \ref{secA1} for mathematical details):
    \begin{itemize}
        \item[(1)] A first solution to the free Hamilton-Jacobi equation is obtained by separation of variables. This first action function reads 
        \begin{equation}\label{1st action function}
            S_1(\bm{q},t)=-\frac{\bm{P}^2}{2m}t + \bm{P}.\bm{q}
        \end{equation}
        where $\bm{P}$ is the initial momentum vector, whose coordinates $(P_1,P_2,P_3)$ are non-additive integration constants. Assuming that we know the initial momentum vector, we can plug (\ref{1st action function}) into the classical guiding equation (\ref{classical guiding equation}) yielding a velocity field on configuration space. The integral curves of this velocity field describe the possible trajectories for the system
        \begin{equation}\label{classical trajectory}
            \bm{Q}(t)=\frac{\bm{P}}{m}t+\bm{Q}(0),
        \end{equation}
        depending on the unknown initial position $\bm{Q}(0)$ of the system. As $\bm{P}$ is a constant vector, the action function (\ref{1st action function}) generates a set of possible trajectories for the system corresponding to parallel lines of same momentum, but starting from different initial positions (see the figures in \cite{callender_one_2015}). Moreover, specifying the initial position $\bm{Q}(0)=\bm{Q}_0$ for the system along with its action function $(S_1,\bm{Q}_0)$ selects a single trajectory
        \begin{equation}\label{exact classical trajectory}
            \bm{Q}(t)=\frac{\bm{P}}{m}t+\bm{Q}_0,
        \end{equation}
        which is the integral curve of (\ref{classical guiding equation}) passing through $\bm{Q}_0$ at time $t=0$.
        \item[(2)] On the other hand, once we know the exact trajectory of the free particle, we can compute its action function by integrating its Lagrangian along the trajectory. Assuming that the free classical particle moves along the trajectory (\ref{exact classical trajectory}) and integrating the free Lagrangian $L=m\Dot{q}^2/2$ along $\bm{Q}(t)=\frac{\bm{P}}{m}t+\bm{Q}_0$, yields (see appendix \ref{secA1}) the action function
        \begin{equation}\label{2nd action function}
            S_2(\bm{q},t)=\frac{m}{2t}(\bm{q}-\bm{Q}_0)^2.
        \end{equation}
        Hence a classical particle moving along the trajectory $\bm{Q}(t)=\frac{\bm{P}}{m}t+\bm{Q}_0$ can be described by different action functions, namely (\ref{1st action function}) and (\ref{2nd action function}).
    \end{itemize}
This example illustrates what Holland calls ``[t]he nonuniqueness of $S$ for a given mechanical problem" \cite[pp.36]{holland1995quantum}. While (\ref{1st action function}) describes the propagation of plane waves on configuration space, (\ref{2nd action function}) describes the propagation of circular waves emanating from the point $\bm{Q}_0$ \cite{callender_one_2015}. From the guiding equation (\ref{classical guiding equation}), we know that the velocity of the particle is always orthogonal to the level sets of the action function, the two action functions---(\ref{1st action function}) and (\ref{2nd action function})---therefore generate two different flows (i.e. set of possible trajectories) for the particle. However, if we specify the initial position and momentum $(\bm{Q}_0,\bm{P})$ of the system, we select the same trajectory in the two sets.\\

Before turning to the differences between the action function and the wave function, let us first emphasize a true similarity between these two variables. As in BM, the specification of the initial position and the action function of a classical particle, i.e. $(\bm{Q}_0,S)$, uniquely determines the motion of this particle. While Callender writes ``As one sees, the $S$-function doesn’t uniquely determine the motion. This is a huge difference between classical and BM. Its importance cannot be underestimated when comparing the two theories physically." \cite{callender_one_2015}, we would say the opposite. The trajectory of a classical particle does not uniquely determine its action function\footnote{Which is closer to Matarese's formulation ``in classical mechanics, our S is derived from the particle trajectories, while in BM, our particle trajectories are derived
from S." \cite[pp.66]{matarese_metaphysics_2023}.} and the action function uniquely determines the particles trajectory, once we know the initial configuration. Despite being perfectly correct, the case of the second action function (\ref{2nd action function}) in Holland's example can mislead us into thinking that $(\bm{Q}_0,S)$ does not uniquely determine the motion of a classical particle, however this ambiguity can be easily removed.\\
As soon as we know the initial position of a classical particle along with its smooth action function, i.e. $(\bm{Q}_0,S)$, one can immediately compute the gradient of the action function, which, according to the classical guiding equation (\ref{classical guiding equation}), gives us the initial momentum $\bm{P}_0$ of the particle. From $(\bm{Q}_0,S)$ we've thus derived $(\bm{Q}_0,\bm{P}_0)$ which, as everyone knows, uniquely determines the motion of a free classical particle. Hence $(\bm{Q}_0,S)$ uniquely determines the motion of a classical particle. So, why would Holland's example mislead us into thinking the contrary?\\
Crucially, starting from $(\bm{Q}_0,S)$, we would be unable to apply the previous reasoning if, for some mathematical reasons, we were unable to compute the gradient of the action function near the initial configuration $\bm{Q}_0$. The second action function (\ref{2nd action function}) perfectly illustrates this point. Indeed, if the initial momentum $\bm{P}_0$ is unknown, $S_2$ (\ref{2nd action function}) generates a set of possible trajectories, all starting from $\bm{Q}_0$ with different (unknown) initial momentum $\bm{P}_0$. In this specific case, since the gradient of the action function (\ref{2nd action function}) is not mathematically well defined at point $\bm{Q}_0$\footnote{If $\bm{q}$ tends to $\bm{Q}_0$ then $t$ tends to zero and the action function (\ref{2nd action function}) and its gradient become undefined.}, we can not use the classical guiding equation (\ref{equation de guidage Hamilton-Jacobi}) to compute the initial momentum of the particle. Therefore, in this particular case, specifying the initial position and the action function $(\bm{Q}_0, S_2)$ does not allow to select a single trajectory for the system.\\
However we should not be lead astray by this specific example. As a matter of fact, except for pathological cases where the gradient of the action function is undefined at the system's initial configuration, specifying the initial configuration and the action function $\big(\bm{Q}_0,S_0(\bm{q})\big)$ uniquely determines the classical system's trajectory. Far from being a  huge difference, this instead represents a strong similarity between BM and HJ formalism.

\subsection{Differences between $\psi$ and the $S$-function}

Building on the work of Holland and Callender, we can nonetheless identify key differences between BM and the classical HJ formalism. Holland summarizes the situation in this long quote
 \begin{quote}
        The essential difference between the classical and quantum treatments of the motion of a particle may be expressed as follows.\\
        In CM the mechanical problem has a unique solution once the external potential is specified and the initial position and velocity of a particle are given. If one formulates the problem in phase space and solves for the motion using the Hamilton-Jacobi equation, all of the (infinite number of) $S$-functions corresponding to the given initial conditions generate the same motion. The system evolution is independent of the 'state' defined by the classical wavefunction. The $S$-functions are distinguished by the global ensembles they generate (cf. §2.3).\\
        In the quantum theory of motion [i.e. BM], specification of the external potential and the initial position and velocity of a particle is not sufficient to determine the motion uniquely; one must specify in addition the quantum state. A set of particles in classically identical initial states (same $\bm{x}_0$, $\bm{P}_0=\nabla S(\bm{x}_0)$, $V(\bm{x})$) are no longer in identical states as their subsequent motions generically differ in infinitely variable ways. This implies that the quantal $S$-functions are distinguished in a much stronger way than their classical counterparts, at the level of individual ensemble elements.
        \cite[pp.131-132]{holland1995quantum}.
    \end{quote}
As emphasized by Holland's example, in CM the initial conditions $(\bm{Q}_0,\bm{P}_0, S_1)$ and $(\bm{Q}_0,\bm{P}_0, S_2)$ select the same trajectory even though $S_1(\bm{q},t)\ne S_2(\bm{q},t)$. More generally, all $S$-functions compatible with the initial condition $(\bm{Q}_0,\bm{P}_0)$ will result in the same trajectory. What really matters is not the action function itself, but only it's gradient at initial configuration---$\nabla S_c(\bm{Q}_0)$---, as it yields $\bm{P_0}$. In CM, the action's only role is to encode initial momentum. Once we have the same initial configuration $\bm{Q}_0$ (or the same initial probability distributions) and the same external potential $V$, we only need $\nabla S_1(\bm{Q}_0)=\nabla S_2(\bm{Q}_0)$ for generating the same trajectory\footnote{Two different $S$-functions that agree on the initial gradient (momentum) but differ elsewhere still produce the same path. The ensemble of trajectories they describe is the same, so the $S$-functions are only different in how they group trajectories, not in individual outcomes. The $S$-functions can generate different ensembles, but each individual trajectory in those ensembles is determined solely by its initial conditions.}. Hence the action function is not truly involved in the classical dynamics, that's why CM can be formulated without reference to the action function $S_c$.\\
In BM, unlike in CM, $(\bm{Q}_0,\bm{P}_0, S_1)$ and $(\bm{Q}_0,\bm{P}_0, S_2)$ generally select distinct trajectories, as soon as $S_1(\bm{q},t)\ne S_2(\bm{q},t)$. This is because the trajectory in BM depends not only on the phase $S_q$ of the wave function, but also on its amplitude $R_q$. Different actions $S_1$ and $S_2$ typically correspond---through the continuity equation---to different amplitudes $R_1$ and $R_2$, leading to different quantum potentials $U^\psi_1$ and $U^\psi_2$ in the quantum Hamilton-Jacobi equation. Hence, even if $\nabla S_1(\bm{Q}_0)=\nabla S_2(\bm{Q}_0)$, we generally have $\nabla S_1(\bm{Q}(t))\ne \nabla S_1(\bm{Q}(t))$, because generally $U^\psi_1\ne U^\psi_2$\footnote{If two different wavefunctions have the same $S$ gradient at the initial point, then the initial momentum $\bm{P}_0$ is the same. But their future gradients of $S$ might differ, leading to different future momenta. Therefore, starting from the same $q$ and $p$, but different $S$ functions (i.e., different wavefunctions), the trajectories can diverge because the subsequent $S$ (as part of the wavefunction) evolves differently.}. Indeed, even if two Bohmian particles share the same initial $\bm{Q}_0$ and $\nabla S_q(\bm{Q}_0)$, their future motion may differ because of the following feedback loop introduced by the quantum potential: $R_q$ alters $S_q$ in quantum HJ equation, which in turn reshapes $R_q$ in the continuity equation.\\
In a nutshell, the classical action function is entirely reducible to the variables $(\bm{q},\bm{p})$---it contains no more physical information than $(\bm{q},\bm{p})$. As every one knows, classical equations of motion can be entirely rewritten without any reference to the action function (e.g., Hamilton's equations or Newton's equations). Hence the motion of a classical system does not strictly rely on the high dimensional field $S_c$, because only its gradient at initial configuration influences the system's trajectory. In stark contrast, the trajectory of a Bohmian system does rely on the high dimensional field $\psi$ on configuration space\footnote{. For instance, in a double-slit experiment, even if two particles start with the same position and momentum, their trajectories depend on the entire interference pattern of the wavefunction. Different wavefunctions (even if they have the same initial gradient at the starting point) would create different interference patterns, leading to different trajectories. This shows how the quantum $S$-function's global properties influence individual paths.}. The wave function is epistemically irreducible: in BM $(\psi,\bm{Q})$ contains more information than $(\bm{q},\bm{p})$. It comes with no surprise since the complete description of Bohmian system is given by $(\psi,\bm{Q})$, and all the different formulations of the BM make use of the wave function, at least indirectly by coupling $R_q$ and $S_q$ through $U^\psi$\footnote{Even the second-order formulation of the Bohmian dynamics used by David Bohm \cite{bohm1952suggested} indirectly relies on the wave function through the quantum potential (\ref{Quantum potential}).}.\\
Interestingly, some attempts have been made to formulate BM without using the wave function \cite{poirier2010bohmian}\cite{Schiff_2012}. However, these attempts involve very complicated equations in order to express the Bohmian laws of motion, and do not constitute a full reformulation of the Bohmian formalism. Contrary to the casual Bohmian formalism relying on the wave function, this reformulations have not be shown correctly predicting a huge variety of quantum phenomena (e.g., double slit experiment, quantum tunneling, EPR-Bell experiment...), nor have they been successfully applied to other quantum equations as Pauli equation or Dirac equation. To be clear, we do not provide any mathematical proof of the impossibility to formulate the Bohmian dynamics without the wave function. Nevertheless, given that we do not know of any complete and clean reformulation of BM that does not use the wave function, the wave function seems to be a necessary ingredient of any formulation of BM.\\
let us conclude this section by summarizing the key differences\footnote{We label them $P-1$ and $P-2$, because this mathematical differences are going to be the premises of a metaphysical argument, in the next section.} between the role played by the wave function in BM and the role played by the $S$-function in CM:
\begin{itemize}
    \item[(P-1)] In CM different\footnote{By ``different functions", we mean different functional forms---such as (\ref{1st action function}) and (\ref{2nd action function})---for the $S$-function, which is a stronger difference that a mere phase factor difference.} $S$-functions can be associated to the same trajectory---what Holland calls the ``non-uniqueness" of $S_c$---whereas in BM a given trajectory is generally associated to a single wave function---to within a phase factor.
    \item[(P-2)] In CM the $S$-function can be reduced (and eliminated) to the variables $(\bm{q},\bm{p})$, whereas in BM the wave function is a necessary and irreducible variable in the complete description of a system, ``[In BM, one] can’t get a well-posed initial value problem without [the wave function]" \cite{callender_one_2015}.
\end{itemize}
Now, what metaphysical lesson should we draw from these differences?

\section{From Dynamical Necessity to Ontological Commitment}\label{sec8}

The failed attempt to provide the wave function with a well-suited classical analog yields several metaphysical lessons. First, it offers an argument in favor of a realist interpretation of the wave function. Second, as we will see in the next section, it supports a strong case for the \textit{sui generis} interpretation.

\subsection{The Causal Agent Argument}

It has already been suggested in the literature that the differences between the quantum wave function and the classical action demand distinct metaphysical treatments of these entities.\\
More specifically, Holland concludes from his example that the quantum wave function is a ``causal agent", while the classical action is not \cite{holland1995quantum}. Callender emphasizes that Holland ``doesn’t mean anything philosophically subtle [by causal agent], but rather merely that something is a causal agent if it’s needed to generate the motion of the beables" \cite{callender_one_2015}. Holland seems nevertheless to be committed to a realist interpretation of the wave function, as he writes ``[o]ntologically the wave and particle are on an equal footing (i.e., they both objectively exist)" \cite[pp.79]{holland1995quantum}. In the same page of his book, Holland describes the wave function of an electron as a `massive' and `charged' field\footnote{``The phrase `an electron of mass $m$' is therefore to be interpreted that both $\psi$ and the corpuscle are associated with the parameter $m$; $\psi$ may be said to be 'massive' (and 'charged' etc.)." \cite[pp.79]{holland1995quantum}.}, strongly suggesting a realist interpretation of the wave function. However, as pointed out by Callender, Holland does not seek a precise ontological definition of the wave function. Without going so far as to describe the wave function as a massive and charged field, we can nevertheless draw a sound argument for \textit{wave function realism}, from Holland's work. Let us see how Callender frames this realist inference, which he calls the \textit{causal agent argument} \cite{callender_one_2015}.\\
First Callender notes:
    \begin{quote}
        This observation seems relevant to the current investigation. The suggestion was that the classical $S$-function is not part of the ontology but is instead part of the nomological structure. That inference seems fine, and even bolstered, by what we’ve learned, namely, that $S$ doesn’t determine the beables’ motion. But the further suggestion that the $S$-function in the quantum case should be treated like the $S$-function in the classical case now seems deeply problematic. One $S$-function is there for convenience, the other by necessity. While admitting that there are no hard and fast philosophical rules in play, that sounds like a relevant difference, one demanding different interpretations of the classical and Bohmian $S$-functions.
        \cite{callender_one_2015}
    \end{quote}
He then concludes:
\begin{quote}
    Putting classical mechanics and BM in the same formalism allows us to appreciate the stark differences between the two, differences that seem relevant to whether ontology stands behind their respective wavefunctions or not. It is therefore perhaps fair to conclude that the natural or even default interpretation of the wavefunction for a Bohmian is that it is ontological.
    \cite{callender_one_2015}
\end{quote}
As discussed in section \ref{sec7}, Holland's example highlights mathematical differences between the classical action function and the quantum wave function. These differences---summarized in the conclusion of section \ref{sec7}---form the premises of the \textit{causal agent argument}, which concludes:
\begin{itemize}
     \item[(C-1)] The classical guiding equation---$\dot{\bm{Q}_k}(t)=\bm{\nabla}_k S_c\big(\mathbf{Q}(t),t\big)/m$---does not describe a causal interaction, whereas the Bohmian guiding equation---$\dot{\bm{Q}_k}(t)=\bm{\nabla}_k S_q\big(\mathbf{Q}(t),t\big)/m$---describes a causal action of the wave function on the particles.
    \item[(C-2)] While $S_{class}$ is purely epistemic, $S_q$ relates to a substantial physical object denoted by $\psi$.
\end{itemize}
Logically, this argument infers a metaphysical conclusion from mathematical premises. It does so by implicitly relying on the intuition that if a mathematical variable is necessary to uniquely determine the motion of particles, then the physical referent of this variable must causally act on those particles. Similarly, if a mathematical variable is necessary for a complete description of the physical state of a system, it must, in some sense, refer to a concrete physical object.\\
To be fair, the epistemic\footnote{We can characterize the necessity of the wave function for having a well-posed initial value problem in BM as an epistemic necessity, in the sense that we do not know the precise physical state of a Bohmian system---and consequently its motion---if we do not know its wave function.} necessity of the wave function in BM does not logically compel us to accept that it causally acts on Bohmian particles, nor that it represents a substantial physical object. Yet, it does provide metaphysical clues about the physical meaning of the wave function. More specifically, the fact that $\psi$ is part of the initial conditions of a Cauchy problem in BM, and evolves according to its own partial differential equation, strongly points toward a realist interpretation.\\
The analogy with the classical action function highlights that the classical action $S_c$ is not coupled to a classical amplitude $R_c$, and thus is not part of a single mathematical object embedding this coupling. In stark contrast, the Madelung decomposition (as performed in Section \ref{sec6.2}) shows that the wave function refers to a single physical entity, coupling dynamical ($S_q$) and probabilistic information ($R_q$) about the Bohmian system. Furthermore, the fact that the $S$-function is not necessary to describe classical motion, while the wave function is necessary to describe particle motion in BM, strongly suggests that there would be no motion in BM without the wave function.  In BM, once the initial configuration $\bm{Q}_0$ specified, no variable can replace the wave function for calculating its time evolution (i.e. solving the guiding equation). Ultimately, the absence of a high-dimensional physical wave guiding classical particles explains why there is no interference pattern and wave-like behavior in CM.

\subsection{Addressing Callender's Objection}

In contrast to the position outlined above, Callender addresses the \textit{causal agent argument} in order to avoid reifying the wave function:
\begin{quote}
    Suppose one is systematizing Bohmian particles with position. Without an $S$-function, we lack a well-posed initial value problem, and therefore potential strength and power. We can’t tell where such a particle will go without this $S$-function. Does that mean the $S$-function is a beable? No, no more so than requiring mass to get a well-posed value problem classically demands that Humeans treat mass as part of the fundamental furniture of the world. Or perhaps a better analogy in the present case, a Humean might justify the postulation of forces as a way of getting the best systematization of the beables without treating forces as themselves beables. Knowing this, the Humean may not be moved by the above “causal agent” argument.
    \cite{callender_one_2015}
\end{quote}
Callender suggests that within a Humean framework, the wave function in BM could be treated as epistemic\footnote{Callender clarifies that Humeans view laws as summaries rather than fundamental entities: ``Humean views, by contrast, understand laws as a particularly powerful summaries of the Its, but not themselves Its. Hence for the Humean they are a special kind of Bit." \cite{callender_one_2015}}---a tool for systematizing patterns in the Humean mosaic, rather than something physical. Indeed, for Humeans, laws of nature are not intrinsic features of reality but epistemic constructs derived from patterns within the ``Humean mosaic"---David Lewis's term for the totality of particular, local facts (particle's trajectories in the case of BM) that constitute the ontological bedrock of reality. For Humeans like Lewis, laws are not ontologically primitive, instead, they emerge a posteriori from the ``Best System Account" (BSA) which identifies laws as the simplest and most informative systematization of the mosaic's regularities. Thus, laws reflect our description of the world’s contingent order, not a metaphysically necessary structure governing it. This renders laws epistemic tools for prediction and explanation, rather than constituents of the world’s fundamental ontology (see \cite{sep-lewis-metaphysics} for a comprehensive introduction, and see \cite{maudlin2007metaphysics} for a critics).\\
The Humean interpretation of BM---sometimes called ``Bohumianism" \cite{miller2014quantum}---is articulated by Esfeld, Lazarovici, Hubert and D\"urr:
\begin{quote}
    Humeanism about laws is applicable to Bohmian mechanics. Assume that one knows the positions of all the particles in the universe throughout the whole history of the universe. The wave-function of the universe then is that description of the universe that achieves, at the end of the universe, the best balance between logical simplicity and empirical content. In other words, the wave-function of the universe supervenes on the distribution of the particles’ positions throughout the whole of space-time; the same goes for the law of motion.
    \cite{esfeld2014ontology}
\end{quote}
Hence, when applied to BM, Lewis's Humeanism argues that only particles and their trajectories are real. The wave function, guiding equation, and Schr\"odinger equation are not fundamental entities but axioms of the BSA---tools that best systematize the mosaic’s patterns.\\
To explain how this philosophical framework addresses the \textit{causal agent argument}, Callender draws an analogy to classical physics: Just as mass is required for a well-posed classical initial value problem (but not considered a beable), the wave function might similarly be part of the ``best system" of laws for systematizing Bohmian particle trajectories. Similarly, forces in CM (e.g., gravitational or electric forces) are often treated as useful tools for predicting particle motion without being fundamental beables. Based on this analogy, Callender argues that Humeans need not treat the wave function as a beable.\\

However, contrary to Callender's claims, we are going to argue that:
\begin{itemize}
    \item[(1)] The wave function is disanaloguous to classical parameters like mass and forces in CM,
    \item[(2)] Humeanism appears as a very unnatural metaphysical interpretation of the Bohmian formalism, inverting its mathematical and explanatory structure.
\end{itemize}
First of all, mass is a constant parameter in CM, not a dynamically evolving variable requiring contingent initial conditions. By contrast, $\psi$ (including its phase $S$) solves the time-dependent Schr\"odinger equation (a PDE) and requires contingent initial conditions $\psi_0(\bm{q})$. This makes $\psi$ an independent dynamical entity, not a fixed parameter. Similarly, forces in Newtonian mechanics are predefined functions of particle positions (e.g., $F=-kx$), derived from instantaneous configurations (e.g., Newtonian gravity and Coulomb electrostatic forces). They are not independent dynamical variables. In stark contrast, $\psi$ evolves independently via its own PDE, and is irreducible to particle configurations. Let us be more concrete about the mathematical structure of BM.\\
Based on the knowledge of some specific initial conditions $\big(\psi_0(\bm{q}),\bm{Q}_0\big)$, BM’s formalism involves two steps for calculating trajectories: (1) solving the Schr\"odinger equation (PDE) for $\psi_t$, given $\psi_0(\bm{q})$; (2) using this solution $\psi_t$ to solve the guiding equation (ODE) for $\bm{Q}(t)$, given $\bm{Q}_0$. This reveals a critical asymmetry: while we need $\psi$ to compute trajectories $\bm{Q}(t)$, $\psi$ evolves independently of $\bm{Q}(t)$. Indeed, in standard quantum mechanics particles have no position, yet we can compute the wave function evolution in the same way than in BM---by solving Schr\"odinger equation. As noted by DGZ\footnote{DGZ use this fact to emphasize the absence of reaction of the particles on the wave function, in BM \cite{durr_bohmian_1995}.}, this demonstrates that particles play no role in the evolution of the wave function. Thus, asserting that $\psi$ supervenes on particle trajectories inverts BM’s mathematical structure: $\psi$ is not constructed from $\bm{Q}(t)$ but constrains it. Humeans might argue that $\Psi_{Un.}$ supervenes holistically on the entire mosaic (all trajectories across spacetime), but this does not resolve the asymmetry: $\psi$ is presupposed in BM’s formalism to define the trajectories, not derived from them.\\

Moreover, if $\psi$ supervened on particle trajectories, this relationship would naturally manifest in the mathematical formalism as a functional dependence: the universal wave function would become a functional of the universal trajectory $Q(t)$\footnote{Relativistically, $t$ should be replaced by an arbitrary real parameter $\lambda$, but this does not affect the current discussion.}, something like $\Psi_{Un.}\big[\bm{Q}(t)\big]$. The natural mathematical operation associated with the Bohumian's supervenience claim would then be as follows: starting from the realized universal trajectory $C:=Q(t)$---a single curve in the universal configuration space---we should, in principle, be able to reconstruct $\Psi_{Un.}$ by integrating the Bohmian dynamics along $C$.\\
Maudlin and Albert have rightly emphasized that ``the complete specification of particle trajectories [...] seems to postulate much less information than the wave function. This is because the particle trajectories form a single curve in configuration space, while the wave function assigns values to every point in configuration space" \cite{chen2019realism}. However, the situation is worse than that. Indeed, the information based solely on the realized Bohmian trajectory is, in general, insufficient to reconstruct or deduce the wave function, even along the single trajectory $C$---let alone on the whole configuration space. This situation contrasts sharply with classical mechanics, where the realized trajectory contains all the necessary information to reconstruct the classical action $S(\bm{q},t)$.\\
To see this more concretely, consider the decomposition of the Bohmian dynamics. The wave function $\Psi_{Un.}(\bm{q},t)$ can theoretically be reconstructed along a trajectory $C$ by integrating the following relations:
\begin{eqnarray*}
\frac{d}{dt}S_q(\mathbf{q},t)=\sum_{i=1}^{N}\frac{m\mathbf{v}_{q,i}^2}{2}-V(\mathbf{q},t)-U^\Psi(\mathbf{q},t)\nonumber\\
\frac{1}{ R_q(\mathbf{q},t)}\frac{d}{dt}R_q(\mathbf{q},t)=-\frac{1}{2}\sum_{i=1}^{N}\boldsymbol{\nabla}_i\cdot\mathbf{v}_{q,i}\nonumber\\
\mathbf{v}_{q,i}=\frac{1}{ m}\boldsymbol{\nabla}_i S_q(\mathbf{q},t)
\end{eqnarray*}
where $\frac{d}{dt}$ denotes the total derivative along $C$.\\
However, these formulas presuppose that the quantities $\nabla^2 R_q$ and $\nabla^2 S_q$ are known along $C$. The mathematical operation $\nabla$ (and by extension $\nabla^2$) involves derivatives not only in the direction tangent to the velocity vector along $C$, but also in directions perpendicular to $C$. This in fact implies knowledge of information associated with trajectories infinitely close to $C$, but currently unrealized.   Mathematically, this poses no problem, as we can assume an infinite number of points in the vicinity of the actual trajectory (in the case of numerical simulation, we can also approach this ideal case with a large number of points in the vicinity of $C$ \cite{deckert22}). However, if we consider the whole universe with a single Bohmian trajectory---as Bohumians do---it is generally impossible to reconstruct $\Psi_{Un.}(\mathbf{q},t)$, even along $C$.\\
This contrasts with classical mechanics, where the absence of the quantum potential and the lack of a conservation law for the amplitude $R_q$ eliminate this issue. In classical mechanics, it is indeed possible to reconstruct $S(\bm{q},t)$ along the realized trajectory without requiring information from neighboring trajectories.\\
Hence the mathematical structure of BM strongly suggests that $\psi$ is not derived from the mosaic---it constrains the mosaic.\\

The Bohumian might argue that the inversion of the mathematical structure is justified because we lack direct empirical access to the wave function, whereas particle positions are measurable. This aligns with Bell's observation that the wave function is `more hidden' than particle positions. However, this argument overlooks a critical point: we also lack direct empirical access to Bohmian trajectories. In fact, Bohmian trajectories are theoretical constructs inferred from the wave function and the guiding equation, not directly observable quantities. Furthermore, one might object that particle positions are at least indirectly measurable through experimental outcomes, such as detector clicks or interference patterns. However, these measurements do not reveal the actual trajectories of particles but rather statistical distributions that depend on the wave function. Thus, the wave function remains indispensable for explaining and predicting these outcomes, further challenging the supervenience claim.\\
In summary, the Bohumian justification for inverting the mathematical structure of BM fails to hold up under scrutiny. The lack of empirical access to Bohmian trajectories, combined with the indispensable role of the wave function in explaining experimental outcomes, makes the supervenience claim highly implausible.\\

Correspondingly, \textit{Bohumianism} inverts the explanatory structure of BM. In standard presentations of BM, particle trajectories are explained via the wave function’s dynamical evolution. Consider the double-slit experiment. When both slits are open, the wave function propagates through both slits, guiding the particle along a trajectory that passes through one slit (e.g., the upper slit). If we close one slit (e.g., the lower slit), the wave function no longer propagates through the closed slit, altering its evolution and thereby modifying the particle’s trajectory. BM thus explains both individual trajectories and collective interference patterns by appealing to the wave function’s spatiotemporal evolution. In stark contrast, \textit{Bohumianism} inverts this explanatory structure. For the Bohumian, the wave function is not a physical entity that produces trajectories but a post hoc construct derived from the totality of particle positions (the Humean mosaic) via the BSA. This raises critical questions: If the wave function is merely a BSA-derived summary of observed trajectories (e.g., interference patterns), how can it simultaneously explain those patterns? Given that in \textit{Bohumianism} the interference pattern partly defines the wave function, either the explanation is circular or there is no explanation for the interference pattern in the double slit experiment. The second option seems to be more likely, for we don't see how some axiom of the BSA could produce interference patterns on the screen. One should note that it is just an instance of the more general claim that Lewis's Humeanism appears to reverse scientific practice
\begin{quote}
    There is a common objection from physics[...]: on Humeanism, the laws of fundamental physics do not have any explanatory function. They sum up, at the end of the universe, what has happened in the universe; but they do not answer the question why what has happened did in fact happen, given certain initial conditions
    \cite{esfeld2014ontology}
\end{quote}
Whereas scientific practice typically derives empirical facts from laws (e.g., predicting trajectories via $\psi$), Humeanism derives laws from empirical facts (the mosaic). For Humeans, metaphysics takes precedence: the ontological primacy of the mosaic justifies treating laws---and $\psi$---as descriptive tools rather than causal agents or governors. Critics argue this clashes with physics’ explanatory norms, where laws are not mere summaries but grounds for prediction and understanding (Davide Romano, p.c.).\\

In summary, the \textit{causal agent argument} provides strong support for a realist interpretation of the wave function in BM. While Callender's Humean objection offers an alternative perspective, it fails to account for the mathematical and explanatory asymmetries inherent in the Bohmian formalism. These considerations suggest that the wave function is best understood as a fundamental ontological entity, rather than a mere epistemic tool.\\

In the next subsection we will preemptively address a potential objection to the \textit{causal agent argument}.

\subsection{Addressing Ontological Discontinuities}

Giving different metaphysical interpretations of the same guiding equation $\bm{v}=\bm{\nabla}S/m$ in the classical regime and in the quantum mechanical regime introduces an ontological discontinuity\footnote{We borrow the term ``ontological discontinuity" to Andrea Oldofredi \cite{Oldofredi2022}.} at the classical limit. From a mathematical perspective, when one goes from quantum mechanical regime to classical regime, the quantum potential vanishes progressively from the quantum Hamilton-Jacobi equation \cite[pp.82]{durr2020understanding}. Consequently the coupling between the amplitude and the phase of the quantum wave function decreases smoothly in (\ref{Decomposed Bohmian dynamics}), such that we end up with the same dynamical equations as the classical dynamics (\ref{Probabilistic HJ}), and with
\begin{equation*}
    S_{class}\approx S_{q}.
\end{equation*}
However if $S_{class}$ does not describe any physical object while $S_q$ does, this raises the question of what this equality means physically. While the classical limit can be described by a smooth mathematical transition from equations (\ref{Decomposed Bohmian dynamics}) to (\ref{Probabilistic HJ}), the physical meaning of the guiding equation changed qualitatively during this process, switching from a causal interaction in the quantum regime to a non-causal interaction in the classical regime. Correspondingly, the physical meaning of the $S$-function experiences an ontological leap, switching from the description of a concrete physical object in the quantum regime to a nomological entity in the classical regime. This suggests that qualitative ontological changes could arise from a quantitative change in the quantum system (i.e., adding new particles to the quantum system).\\
Ironically, one should note that the same kind of metaphysical leap (or ontological discontinuity) may occur in the nomological interpretation of the wave function (despite not being brought by logical necessity in this case). When describing the double slit experiment, many Bohmians explain the interference pattern by the propagation of a wave. As the particles are sent one by one through the slits, their wave function is defined on a $3$-dimensional configuration space. The dimensional mismatch between configuration space and the physical space thus vanishes, and it becomes tempting to explain the experimental outcomes of the double slit experiment by the propagation of a wave in physical space. Here's for instance what Sheldon Goldstein writes about the double slit experiment
\begin{quote}
    While each trajectory passes through only one slit, the wave passes through both; the interference profile that therefore develops in the wave generates a similar pattern in the trajectories guided by the wave. \cite{sep-qm-bohm}.
\end{quote}
There's no question that this explanation of the seemingly paradoxical results of the double-slit experiment sounds really convincing to any Bohmian. However it requires a \textit{wave-and-particle} ontology. As a matter of fact, if the wave function is not ``an element of the physical reality", as DGZ claim \cite[pp.95]{ney_wave_2013}, how could it propagate through the slits and create interference patterns? Hence for this explanation to hold, the wave function has to be interpreted as describing a real wave propagating in physical space. That's probably why every Bohmian tends to be a wave function realist when introducing BM in the context of the double slit experiment. However as soon as we add a particle and consider a 2-particle system, because the wave function is now defined on a six dimensional configuration space, proponents of the nomological interpretation change their ontology: the wave function is no longer part of the physical reality.\\
Since there is no compelling reason to qualitatively change the ontology of a physical theory based on the number of particles in a system, maintaining the nomological interpretation of the wave function would require abandoning the causal and highly convincing Bohmian explanation of the double-slit experiment. In other words, the nomological interpretation of the wave function appears incompatible with the very argument that initially persuaded most proponents of BM. Regardless of whether we adopt a realist or nomological interpretation of the wave function, our ontology should remain consistent across systems of varying particle numbers—qualitative ontological changes should not arise from mere quantitative differences. For the nomological interpretation, resolving this ontological discontinuity would necessitate giving up the compelling Bohmian explanation of the double-slit experiment, which seems a high price to pay.\\
As far as the realist interpretation of the wave function is concerned, the ontological discontinuity between $S_q$ and $S_c$ cannot be canceled out. However, such an ontological discontinuity seems less troublesome when it occurs at the boundary between two different theories, namely BM and CM, than an ontological discontinuity within a single theory. As explained by Oldofredi a similar ontological discontinuity also occurs at the boundary between Bell-type quantum field theory and CM. As a matter of fact, Bell-type QFT postulates fermion number density as local beables, yet ``[fermion number density] have no classical analogues, implying an ontological discontinuity between the classical and the quantum regime"\cite{Oldofredi2022}. Oldofredi goes on pointing out that such a discontinuity was tolerated by Bell, probably because of the provisional character of physical theories. In any case one may also simply argue that CM is actually wrong, and therefore it is unsurprising that its ontology does not align with that of quantum mechanics. This is particularly true when considering the wave function, which introduces a fundamentally new feature in the quantum formalism.\\

In a nutshell, we've seen in this section that the epistemic necessity of the wave function for generating the motion of the Bohmian particles, points in the direction of a realist interpretation of the wave function. Admittedly, treating the action function and the wave function differently introduces a metaphysical discontinuity at the classical limit. However, this can be mitigated by recognizing that CM is ultimately incorrect or, at best, incomplete as a fundamental physical theory. On the other hand, the nomological interpretation of the wave function introduces an ontological discontinuity within BM itself. Furthermore, the cost of resolving this ontological discontinuity is the loss of the Bohmian explanation that initially convinces most proponents of the theory.\\
Lastly, the epistemic differences between the wave function and the classical action function reveal that our initial attempt to provide the wave function with a convincing classical analogue ultimately fails, which points toward a \textit{sui generis} interpretation of the wave function.

\section{The Wave Function’s Ontological Novelty: Toward a \textit{Sui Generis} Interpretation}\label{sec9}

Our journey from Hamiltonian to $S$-function analogies exemplifies a dialectic of failed reductions. Each classical counterpart ultimately highlights the wave function’s ontological uniqueness.
More specifically, the ontological novelty of the wave function in BM is underscored by two interrelated observations:
\begin{itemize}
    \item[(1)] \textbf{The Ad-Hoc Invention of ``Nomological Entity" as a Conceptual Placeholder:} To our knowledge the term ``nomological entity" emerged \textit{ad hoc} in the debate to resolve ambiguities about the physical meaning of the wave function---neither fully law-like nor a classical field. This terminological innovation itself signals the wave function’s ontological novelty: it resists categorization within existing metaphysical frameworks (e.g., physical fields, laws). By retroactively labeling classical entities as the Hamiltonian as 'nomological', proponents of the nomological interpretation aim to (a) provide physicists and philosophers with some metaphysical intuitions about this new (and thus unfamiliar) notion of nomological entity and (b) domesticate the wave function's novelty, framing it as familiar rather than revolutionary. Roughly speaking, if the wave function is a nomological entity, just as the classical Hamiltonian, then there is nothing particularly strange about it.
    \item[(2)] \textbf{Failure of Classical Analogues:} While classical entities, like the Hamiltonian, are retroactively labeled 'nomological', none replicate the wave function’s singular role in BM, namely a dynamical variable solution of a wave equation, necessary for the theory's formulation, generating a deterministic particle dynamics while encoding a probability distribution.   
\end{itemize}
 The first point demonstrates that the very invention of the notion ``nomological entity" already sounds as a conceptual placeholder for the wave function's \textit{sui generis} status. The second points strengthens this conclusion: even if we were to assume that the wave function is nomological, it would nevertheless represent an entirely new type of nomological entity, a nomological entity of its own kind. In other words, the inability to find a classical nomological entity with dynamical, epistemic and probabilistic features matching those of the wave function leaves us with an unfamiliar metaphysical category---nomological entity---created specifically for the wave function and occupied solely by it. This observation renders the “nomological” label redundant and necessitate direct acknowledgment of the wave function’s \textit{sui generis} ontological status. That is to say, the wave function in fact represents a new kind of physical entity, or what Maudlin calls a \textit{Quantum State}:
\begin{quote}
     We do not even know the right general ontological category in which to put it. Indeed, there is no reason to believe that any theorizing or speculation on the nature of the physical world that took place before the advent of quantum theory would have hit on the right ontological category for the quantum state: because it is so hidden, there would have been nothing relevant to speculate about. Whether one finds the possibility invigorating or disheartening, the best ontological category for the quantum state might simply be the category \textit{Quantum State}, just as the right ontological category for a classical field is Field, not “stress in a medium” or “collection of particles. \cite[pp.152]{maudlin2013nature}
\end{quote}
This amounts to a \textit{sui generis} interpretation of the wave function \cite{chen2019realism}, as the physical entity described by the wave function---a quantum state---can not be compared to any other preexisting physical entity.
\begin{quote}
    The quantum state is a novel feature of reality on any view,
    and there is nothing wrong with allowing it a novel category:
    quantum state. This is, of course, not an informative thing to say,
    but it does free us from the misguided desire to liken the quantum state to anything we are already familiar with. \cite[pp.89]{maudlin2019philosophy}
\end{quote}
Given the novelty of the physical entity known as the \textit{Quantum State}, it may seem questionable to impose metaphysical requirements derived from classical physics, such as the action-reaction principle. Instead we must accept the novel physical features of the \textit{Quantum State} and make it part of our physics. Its most distinctive feature lies in the fact that it is a non-local beable. While the \textit{Primitive Ontology} approach restricts its ontology to local beables exclusively, the \textit{sui generis} interpretation of the wave function introduces a non-local beable into the Bohmian ontology.\\
When interpreting the physical meaning wave function in the context of BM, two important facts must be considered. First, there is a striking dimensional mismatch between the 3N-dimensional configuration space and the 3-dimensional physical space. Second, the complete physical description of any Bohmian system is given by its configuration and wave function $(\bm{Q},\psi)$, not by its configuration alone, and $\psi$ cannot be derived from other primitive variables. From our perspective, this straightforward epistemic fact makes it highly unlikely that $\psi$ does not represent some feature or property of matter, let alone that it is excluded from physical reality. Considering the Bohmian ontology, we face a choice between two conflicting principles\footnote{This analysis is close to Valia Allori's quote: ``The primitive ontology provides us with a clear metaphysical picture of the
world. So does the wave function ontology: the world is made of stuff represented by the wave function. One difference between the approaches is that the primitive ontology is in three-dimensional space (or in space-time), whereas the wave function is not. As a consequence of this, in the case of the wave function ontology, the scientific image does not have much in common with the previously accepted Newtonian picture. Th is is not true in the case of theories with a primitive ontology. In contrast to the case of wave function ontology, the primitive ontology approach reflects the desire to keep the scientific image closer to the classical way of understanding things, given that it is possible."\cite[pp.62]{allori2013primitive}.}:
\begin{itemize}
    \item[(1)] As prescribed by the primitivist approach, matter can only be represented by 3-dimensional mathematical variables.
    \item[(2)] Based on an epistemic criterion, any mathematical variable necessary to fully describe the physical state of a system should be part of the theory's primitive ontology.
\end{itemize}
While the first principle seems reasonable, introducing non-local beables is not unreasonable, especially given the inherent non-locality of quantum mechanics. Moreover, the Bohmian laws of motion are already defined by two elegant equations, which, unlike the wave function, possess the appropriate mathematical structure to represent laws of nature. So, why should we add a metaphysically ambiguous nomological entity to the Bohmian ontology?\footnote{An alternative to abandoning the nomological interpretation would be to consider BM as an incomplete theory. Within the framework of de Broglie's double solution \cite{de1958tentative}, BM would be an approximation of a more sophisticated theory in which both the material point and the wave function emerge as effective variables describing the quantum system. In this framework, the fact that the wave function is a non-local beable in BM would not be problematic, as at a more fundamental, `subquantum' level, the quantum system would be described by new hidden variables defined in space-time, aligning with the intuition behind the primitivist approach. However, this would imply a new physics and thus constitutes a separate research program from BM, even though the two may overlap in a subdomain of approximation.}

\backmatter

\bmhead{Acknowledgements}

We thank Davide Romano, Vincent Lam, Siddhant Das and Jean Bricmont for helpful discussions. The authors gratefully acknowledge the insightful comments and constructive suggestions provided by the two anonymous reviewers. Their feedback significantly improved the quality of this manuscript.

\begin{appendices}

\section{Explicit Calculations of Holland's Example}\label{secA1}

In this appendix, we explicit the calculations yielding the two different actions functions $S_1$ (\ref{1st action function}) and $S_2$ (\ref{2nd action function}) in the case of a free classical particle. Holland gives the solutions in his book \cite[pp.36]{holland1995quantum} and Callender builds on them in his work \cite{callender_one_2015}. However, to our best knowledge, the calculations have not been made explicit in any book or paper. Even if these calculations can be regarded as trivial, we feel it is useful to make them explicit, since our argument is based on their results.\\
In \cite[pp.36]{holland1995quantum}, Holland takes the example of the free Hamilton-Jacobi equation\footnote{To simplify the reasoning let us consider the case of a one dimensional free classical particle, whose single spatial degree of freedom is denoted $q$.}
\begin{equation}\label{Free HJ}
    \frac{\partial S(q,t)}{\partial t}+\frac{(\nabla S(q,t))^2}{2m}=0
\end{equation}
To show that the action function is multi-valued for a given mechanical problem. His proof consists in two steps.\\
He first solves this equation by using separation of variables. To be more specific, assuming that $S(q,t)$ can be written as the sum of a spatial part $W(q)$ and a temporal part $T(t)$, we express the action function as follow
\begin{equation}\label{separation of variables}
    S(q,t)=W(q)+T(t).
\end{equation}
Plugging this expression into (\ref{Free HJ}) yields
\begin{equation}\label{new HJ}
    \frac{\partial T(t)}{\partial t}+\frac{(\nabla W(q))^2}{2m}=0,
\end{equation}
that is to say
\begin{equation*}
    -\frac{\partial T(t)}{\partial t}=\frac{(\nabla W(q))^2}{2m}.
\end{equation*}
Since each side of this equation depends on different independent variables, both sides are necessary equal to the same constant, let's call it $E$. Therefore we have
\begin{equation*}
    \frac{\partial T(t)}{\partial t}=-E,\quad \frac{(\nabla W(q))^2}{2m}=E.
\end{equation*}
Since $S(q,t)=W(q)+T(t)$, we have $\partial_t S(q,t)=\partial_t T(t)$, thus
\begin{equation}\label{partie temporelle}
    \frac{\partial S(q,t)}{\partial t}=-E.
\end{equation}
As the action is the integral over time of the Lagrangian, $\partial_t S(q,t)$ is an energy, thus the constant $E$ denotes the energy of the free particle. Since it is a free particle we have $E=P^2/2m$, $P$ being the constant momentum of the free particle. Plugging (\ref{partie temporelle}) into (\ref{new HJ}) we have
\begin{equation*}
    \frac{(\nabla W(q))^2}{2m}=\frac{P^2}{2m},
\end{equation*}
yielding the ODE\footnote{This describes a spatial wavefront whose gradient corresponds to the momentum $P$.}
\begin{equation}\label{ODE HJ}
    \nabla W(q)=P.
\end{equation}
Recalling that $P$ is a constant, the exact solution to (\ref{ODE HJ}) reads
\begin{equation}\label{solution 1}
    W(q)=P.q+W(0).
\end{equation}
On the other hand the first ODE $\frac{\partial T(t)}{\partial t}=-E$ is easily solved
\begin{equation}\label{solution 2}
    T(t)=-\frac{P^2}{2m}t+T(0).
\end{equation}
Plugging our two solutions (\ref{solution 1}) and (\ref{solution 2}) into (\ref{separation of variables}), we end up with the action function
\begin{equation}\label{first action}
    S(q,t)=P.q-\frac{P^2}{2m}t+S_0,
\end{equation}
with the constant of integration $S_0:=S(0,0)=W(0)+T(0)$ being the value of $S$ at time $t=0$ and position $q=0$.\\
Now we can compute the trajectory $Q(t)$ of this particle. Using the classical guiding equation $v=\nabla S/m$, we find
\begin{equation*}
    \Dot{Q}(t)=\frac{P}{m},
\end{equation*}
and an exact solution to this equation reads
\begin{equation}\label{first trajectory}
    Q(t)=\frac{P}{m}t+Q(0).
\end{equation}
Holland further explains that we can obtain a different action function $S'$ from this trajectory, by integrating the Lagrangian of the free particle along the trajectory (\ref{first trajectory}). For a free particle the Lagrangian reads $L(q,\Dot{q},t)=m\Dot{q}^2/2$. To integrate this Lagrangian along the trajectory $Q(t)$ with $t\in[0,t_1]$ we write
\begin{equation}\label{integrale curviligne}
    \begin{split}
        S'[Q(t)]&=\int_{0}^{t_1}L(Q(t),\Dot{Q}(t),t)dt\\
        &=\int_{0}^{t_1}\frac{p^2}{2m}dt\\
        &=\frac{p^2}{2m}t_1\\
    \end{split}
\end{equation}
where we used $L(Q(t),\Dot{Q}(t),t)=\frac{p^2}{2m}$. Moreover, since the free particle moves along $Q(t)$ with a constant momentum $p=P$, we have the relation $P=mv=m\frac{(Q(t_1)-Q_0)}{t_1}$. Plugging this into (\ref{integrale curviligne}) yields 
\begin{equation}
    \begin{split}
        S'[Q(t)]&=\bigg(\frac{(Q(t_1)-Q_0)}{t_1}\bigg)^2\frac{t_1}{2m}\\
        &=\frac{m}{2t_1}\big(Q(t_1)-Q_0\big)^2.
    \end{split}
\end{equation}
Relabeling $Q(t_1)=:q$, $t_1=:t$ and $Q_0=:q_0$ we end up with the action function
\begin{equation}
    S'(q,t,q_0,0)=\frac{m}{2t}(q-q_0)^2
\end{equation}
This two action functions have been calculated for a classical particle moving in one dimension, but they can be easily extended to three dimension by replacing $q$ by the vector $\bm{q}=(x,y,z)$ and $p$ by $\bm{p}=(p_x,p_y,p_z)$ (in Cartesian coordinates).

\end{appendices}

\bibliography{sn-bibliography}


\begin{thebibliography}{36}
\ifx \bisbn   \undefined \def \bisbn  #1{ISBN #1}\fi
\ifx \binits  \undefined \def \binits#1{#1}\fi
\ifx \bauthor  \undefined \def \bauthor#1{#1}\fi
\ifx \batitle  \undefined \def \batitle#1{#1}\fi
\ifx \bjtitle  \undefined \def \bjtitle#1{#1}\fi
\ifx \bvolume  \undefined \def \bvolume#1{\textbf{#1}}\fi
\ifx \byear  \undefined \def \byear#1{#1}\fi
\ifx \bissue  \undefined \def \bissue#1{#1}\fi
\ifx \bfpage  \undefined \def \bfpage#1{#1}\fi
\ifx \blpage  \undefined \def \blpage #1{#1}\fi
\ifx \burl  \undefined \def \burl#1{\textsf{#1}}\fi
\ifx \doiurl  \undefined \def \doiurl#1{\url{https://doi.org/#1}}\fi
\ifx \betal  \undefined \def \betal{\textit{et al.}}\fi
\ifx \binstitute  \undefined \def \binstitute#1{#1}\fi
\ifx \binstitutionaled  \undefined \def \binstitutionaled#1{#1}\fi
\ifx \bctitle  \undefined \def \bctitle#1{#1}\fi
\ifx \beditor  \undefined \def \beditor#1{#1}\fi
\ifx \bpublisher  \undefined \def \bpublisher#1{#1}\fi
\ifx \bbtitle  \undefined \def \bbtitle#1{#1}\fi
\ifx \bedition  \undefined \def \bedition#1{#1}\fi
\ifx \bseriesno  \undefined \def \bseriesno#1{#1}\fi
\ifx \blocation  \undefined \def \blocation#1{#1}\fi
\ifx \bsertitle  \undefined \def \bsertitle#1{#1}\fi
\ifx \bsnm \undefined \def \bsnm#1{#1}\fi
\ifx \bsuffix \undefined \def \bsuffix#1{#1}\fi
\ifx \bparticle \undefined \def \bparticle#1{#1}\fi
\ifx \barticle \undefined \def \barticle#1{#1}\fi
\bibcommenthead
\ifx \bconfdate \undefined \def \bconfdate #1{#1}\fi
\ifx \botherref \undefined \def \botherref #1{#1}\fi
\ifx \url \undefined \def \url#1{\textsf{#1}}\fi
\ifx \bchapter \undefined \def \bchapter#1{#1}\fi
\ifx \bbook \undefined \def \bbook#1{#1}\fi
\ifx \bcomment \undefined \def \bcomment#1{#1}\fi
\ifx \oauthor \undefined \def \oauthor#1{#1}\fi
\ifx \citeauthoryear \undefined \def \citeauthoryear#1{#1}\fi
\ifx \endbibitem  \undefined \def \endbibitem {}\fi
\ifx \bconflocation  \undefined \def \bconflocation#1{#1}\fi
\ifx \arxivurl  \undefined \def \arxivurl#1{\textsf{#1}}\fi
\csname PreBibitemsHook\endcsname

\bibitem[\protect\citeauthoryear{Bricmont}{2016}]{bricmont2016making}
\begin{bbook}
\bauthor{\bsnm{Bricmont}, \binits{J.}}:
\bbtitle{Making Sense of Quantum Mechanics}
vol. \bseriesno{37}.
\bpublisher{Springer},
\blocation{Cham}
(\byear{2016})
\end{bbook}
\endbibitem

\bibitem[\protect\citeauthoryear{Durr and Mechanics}{2009}]{durr2009physics}
\begin{botherref}
\oauthor{\bsnm{Durr}, \binits{D.}},
\oauthor{\bsnm{Mechanics}, \binits{S.T.B.}}:
The physics and mathematics of quantum theory.
Fundamental Theories of Physics. Springer Berlin Heidelberg
(2009)
\end{botherref}
\endbibitem

\bibitem[\protect\citeauthoryear{Goldstein}{2024}]{sep-qm-bohm}
\begin{bchapter}
\bauthor{\bsnm{Goldstein}, \binits{S.}}:
\bctitle{{Bohmian Mechanics}}.
In: \beditor{\bsnm{Zalta}, \binits{E.N.}},
\beditor{\bsnm{Nodelman}, \binits{U.}} (eds.)
\bbtitle{The {Stanford} Encyclopedia of Philosophy},
\bedition{{S}ummer 2024} edn.
\bpublisher{Metaphysics Research Lab, Stanford University},
\blocation{Stanford}
(\byear{2024})
\end{bchapter}
\endbibitem

\bibitem[\protect\citeauthoryear{Goldstein and Teufel}{1999}]{goldstein_quantum_1999}
\begin{botherref}
\oauthor{\bsnm{Goldstein}, \binits{S.}},
\oauthor{\bsnm{Teufel}, \binits{S.}}:
Quantum spacetime without observers: ontological clarity and the conceptual foundations of quantum gravity.
arXiv.
arXiv:quant-ph/9902018
(1999).
\url{http://arxiv.org/abs/quant-ph/9902018}
Accessed 2024-05-29
\end{botherref}
\endbibitem

\bibitem[\protect\citeauthoryear{Chen}{2019}]{chen2019realism}
\begin{barticle}
\bauthor{\bsnm{Chen}, \binits{E.K.}}:
\batitle{Realism about the wave function}.
\bjtitle{Philosophy compass}
\bvolume{14}(\bissue{7}),
\bfpage{12611}
(\byear{2019})
\end{barticle}
\endbibitem

\bibitem[\protect\citeauthoryear{Ney and Albert}{2013}]{ney_wave_2013}
\begin{bbook}
\beditor{\bsnm{Ney}, \binits{A.}},
\beditor{\bsnm{Albert}, \binits{D.Z.}} (eds.):
\bbtitle{The Wave Function: Essays on the Metaphysics of Quantum Mechanics}.
\bpublisher{Oxford University Press},
\blocation{Oxford ; New York}
(\byear{2013})
\end{bbook}
\endbibitem

\bibitem[\protect\citeauthoryear{Holland}{1995}]{holland1995quantum}
\begin{bbook}
\bauthor{\bsnm{Holland}, \binits{P.R.}}:
\bbtitle{The Quantum Theory of Motion: an Account of the de Broglie-Bohm Causal Interpretation of Quantum Mechanics}.
\bpublisher{Cambridge university press},
\blocation{Cambridge}
(\byear{1995})
\end{bbook}
\endbibitem

\bibitem[\protect\citeauthoryear{Valentini}{1992}]{valentini1992pilot}
\begin{botherref}
\oauthor{\bsnm{Valentini}, \binits{A.}}:
On the pilot-wave theory of classical, quantum and subquantum physics
(1992)
\end{botherref}
\endbibitem

\bibitem[\protect\citeauthoryear{Matarese}{2023}]{matarese_metaphysics_2023}
\begin{bbook}
\bauthor{\bsnm{Matarese}, \binits{V.}}:
\bbtitle{The {Metaphysics} of {Bohmian} {Mechanics}: {A} {Comprehensive} {Guide} to the {Different} {Interpretations} Of {Bohmian} {Ontology}}.
\bpublisher{De Gruyter},
\blocation{Berlin/Boston}
(\byear{2023}).
\doiurl{10.1515/9783110793871} .
\burl{https://www.degruyter.com/document/doi/10.1515/9783110793871/html}
Accessed 2024-05-23
\end{bbook}
\endbibitem

\bibitem[\protect\citeauthoryear{Bell}{1987}]{bell_speakable_1987}
\begin{bbook}
\bauthor{\bsnm{Bell}, \binits{J.S.}}:
\bbtitle{Speakable and Unspeakable in Quantum Mechanics: Collected Papers on Quantum Philosophy}.
\bpublisher{Cambridge University Press},
\blocation{Cambridge [Cambridgeshire] ; New York}
(\byear{1987})
\end{bbook}
\endbibitem

\bibitem[\protect\citeauthoryear{Dürr et~al.}{1995}]{durr_bohmian_1995}
\begin{botherref}
\oauthor{\bsnm{Dürr}, \binits{D.}},
\oauthor{\bsnm{Goldstein}, \binits{S.}},
\oauthor{\bsnm{Zanghì}, \binits{N.}}:
Bohmian {Mechanics} and the {Meaning} of the {Wave} {Function}.
arXiv.
arXiv:quant-ph/9512031
(1995).
\url{http://arxiv.org/abs/quant-ph/9512031}
Accessed 2023-12-20
\end{botherref}
\endbibitem

\bibitem[\protect\citeauthoryear{Hubert and Romano}{2017}]{hubert_romano}
\begin{botherref}
\oauthor{\bsnm{Hubert}, \binits{M.}},
\oauthor{\bsnm{Romano}, \binits{D.}}:
The wave-function as a multi-field.
European Journal for Philosophy of Science
\textbf{8}
(2017)
\doiurl{10.1007/s13194-017-0198-9}
\end{botherref}
\endbibitem

\bibitem[\protect\citeauthoryear{Goldstein and Zanghì}{2011}]{goldstein_reality_2011}
\begin{botherref}
\oauthor{\bsnm{Goldstein}, \binits{S.}},
\oauthor{\bsnm{Zanghì}, \binits{N.}}:
Reality and the {Role} of the {Wavefunction} in {Quantum} {Theory}.
arXiv.
arXiv:1101.4575 [quant-ph]
(2011).
\url{http://arxiv.org/abs/1101.4575}
Accessed 2024-05-23
\end{botherref}
\endbibitem

\bibitem[\protect\citeauthoryear{Allori}{}]{allori_primitive_nodate}
\begin{botherref}
\oauthor{\bsnm{Allori}, \binits{V.}}:
Primitive {Ontology} in a {Nutshell}
\end{botherref}
\endbibitem

\bibitem[\protect\citeauthoryear{Allori et~al.}{2013}]{allori2013primitive}
\begin{botherref}
\oauthor{\bsnm{Allori}, \binits{V.}}, et al.:
Primitive ontology and the structure of fundamental physical theories.
The wave function: Essays on the metaphysics of quantum mechanics,
58--75
(2013)
\end{botherref}
\endbibitem

\bibitem[\protect\citeauthoryear{D{\"u}rr and Lazarovici}{2020}]{durr2020understanding}
\begin{botherref}
\oauthor{\bsnm{D{\"u}rr}, \binits{D.}},
\oauthor{\bsnm{Lazarovici}, \binits{D.}}:
Understanding quantum mechanics.
Cham: Springer
(2020)
\end{botherref}
\endbibitem

\bibitem[\protect\citeauthoryear{Struyve}{2024}]{struyve_bohmian_2024}
\begin{bchapter}
\bauthor{\bsnm{Struyve}, \binits{W.}}:
\bctitle{The {Bohmian} {Solution} to the {Problem} of {Time}}.
In: \beditor{\bsnm{Bassi}, \binits{A.}},
\beditor{\bsnm{Goldstein}, \binits{S.}},
\beditor{\bsnm{Tumulka}, \binits{R.}},
\beditor{\bsnm{Zanghì}, \binits{N.}} (eds.)
\bbtitle{Physics and the {Nature} of {Reality}: {Essays} in {Memory} of {Detlef} {Dürr}},
pp. \bfpage{203}--\blpage{215}.
\bpublisher{Springer},
\blocation{Cham}
(\byear{2024}).
\doiurl{10.1007/978-3-031-45434-9_15} .
\burl{\url{https://doi.org/10.1007/978-3-031-45434-9_15}}
\end{bchapter}
\endbibitem

\bibitem[\protect\citeauthoryear{Giulini et~al.}{2003}]{giulini2003quantum}
\begin{bbook}
\bauthor{\bsnm{Giulini}, \binits{D.J.}},
\bauthor{\bsnm{Kiefer}, \binits{C.}},
\bauthor{\bsnm{L{\"a}mmerzahl}, \binits{C.}}:
\bbtitle{Quantum Gravity: From Theory to Experimental Search}
vol. \bseriesno{631}.
\bpublisher{Springer},
\blocation{Berlin, Heidelberg}
(\byear{2003})
\end{bbook}
\endbibitem

\bibitem[\protect\citeauthoryear{R.~Brown and Wallace}{2005}]{r2005solving}
\begin{barticle}
\bauthor{\bsnm{R.~Brown}, \binits{H.}},
\bauthor{\bsnm{Wallace}, \binits{D.}}:
\batitle{Solving the measurement problem: De broglie--bohm loses out to everett}.
\bjtitle{Foundations of Physics}
\bvolume{35},
\bfpage{517}--\blpage{540}
(\byear{2005})
\end{barticle}
\endbibitem

\bibitem[\protect\citeauthoryear{Vilenkin}{1998}]{PhysRevD.58.067301}
\begin{barticle}
\bauthor{\bsnm{Vilenkin}, \binits{A.}}:
\batitle{Wave function discord}.
\bjtitle{Phys. Rev. D}
\bvolume{58},
\bfpage{067301}
(\byear{1998})
\doiurl{10.1103/PhysRevD.58.067301}
\end{barticle}
\endbibitem

\bibitem[\protect\citeauthoryear{Callender}{2015}]{callender_one_2015}
\begin{barticle}
\bauthor{\bsnm{Callender}, \binits{C.}}:
\batitle{One world, one beable}.
\bjtitle{Synthese}
\bvolume{192}(\bissue{10}),
\bfpage{3153}--\blpage{3177}
(\byear{2015})
\doiurl{10.1007/s11229-014-0582-3} .
Accessed 2024-03-12
\end{barticle}
\endbibitem

\bibitem[\protect\citeauthoryear{Bohm}{1952}]{bohm1952suggested}
\begin{barticle}
\bauthor{\bsnm{Bohm}, \binits{D.}}:
\batitle{A suggested interpretation of the quantum theory in terms of" hidden" variables. i}.
\bjtitle{Physical review}
\bvolume{85}(\bissue{2}),
\bfpage{166}
(\byear{1952})
\end{barticle}
\endbibitem

\bibitem[\protect\citeauthoryear{Bohm and Hiley}{2006}]{bohm2006undivided}
\begin{bbook}
\bauthor{\bsnm{Bohm}, \binits{D.}},
\bauthor{\bsnm{Hiley}, \binits{B.J.}}:
\bbtitle{The Undivided Universe: An Ontological Interpretation of Quantum Theory}.
\bpublisher{Routledge}, \blocation{???}
(\byear{2006})
\end{bbook}
\endbibitem

\bibitem[\protect\citeauthoryear{Madelung}{1927}]{madelung1927quantum}
\begin{barticle}
\bauthor{\bsnm{Madelung}, \binits{E.}}:
\batitle{Quantum theory in hydrodynamical form}.
\bjtitle{z. Phys}
\bvolume{40},
\bfpage{322}
(\byear{1927})
\end{barticle}
\endbibitem

\bibitem[\protect\citeauthoryear{Mauro}{2003}]{mauro2003topicskoopmanvonneumanntheory}
\begin{botherref}
\oauthor{\bsnm{Mauro}, \binits{D.}}:
Topics in Koopman-von Neumann Theory
(2003).
\url{https://arxiv.org/abs/quant-ph/0301172}
\end{botherref}
\endbibitem

\bibitem[\protect\citeauthoryear{Poirier}{2010}]{poirier2010bohmian}
\begin{barticle}
\bauthor{\bsnm{Poirier}, \binits{B.}}:
\batitle{Bohmian mechanics without pilot waves}.
\bjtitle{Chemical Physics}
\bvolume{370}(\bissue{1-3}),
\bfpage{4}--\blpage{14}
(\byear{2010})
\end{barticle}
\endbibitem

\bibitem[\protect\citeauthoryear{Schiff and Poirier}{2012}]{Schiff_2012}
\begin{botherref}
\oauthor{\bsnm{Schiff}, \binits{J.}},
\oauthor{\bsnm{Poirier}, \binits{B.}}:
Communication: Quantum mechanics without wavefunctions.
The Journal of Chemical Physics
\textbf{136}(3)
(2012)
\doiurl{10.1063/1.3680558}
\end{botherref}
\endbibitem

\bibitem[\protect\citeauthoryear{Hall et~al.}{2024}]{sep-lewis-metaphysics}
\begin{bchapter}
\bauthor{\bsnm{Hall}, \binits{N.}},
\bauthor{\bsnm{Rabern}, \binits{B.}},
\bauthor{\bsnm{Schwarz}, \binits{W.}}:
\bctitle{{David Lewis’s Metaphysics}}.
In: \beditor{\bsnm{Zalta}, \binits{E.N.}},
\beditor{\bsnm{Nodelman}, \binits{U.}} (eds.)
\bbtitle{The {Stanford} Encyclopedia of Philosophy},
\bedition{{S}pring 2024} edn.
\bpublisher{Metaphysics Research Lab, Stanford University}, \blocation{???}
(\byear{2024})
\end{bchapter}
\endbibitem

\bibitem[\protect\citeauthoryear{Maudlin}{2007}]{maudlin2007metaphysics}
\begin{bbook}
\bauthor{\bsnm{Maudlin}, \binits{T.}}:
\bbtitle{The Metaphysics Within Physics}.
\bpublisher{Oxford University Press},
\blocation{New York}
(\byear{2007})
\end{bbook}
\endbibitem

\bibitem[\protect\citeauthoryear{Miller}{2014}]{miller2014quantum}
\begin{barticle}
\bauthor{\bsnm{Miller}, \binits{E.}}:
\batitle{Quantum entanglement, bohmian mechanics, and humean supervenience}.
\bjtitle{Australasian Journal of Philosophy}
\bvolume{92}(\bissue{3}),
\bfpage{567}--\blpage{583}
(\byear{2014})
\end{barticle}
\endbibitem

\bibitem[\protect\citeauthoryear{Esfeld et~al.}{2014}]{esfeld2014ontology}
\begin{botherref}
\oauthor{\bsnm{Esfeld}, \binits{M.}},
\oauthor{\bsnm{Hubert}, \binits{M.}},
\oauthor{\bsnm{Lazarovici}, \binits{D.}},
\oauthor{\bsnm{D{\"u}rr}, \binits{D.}}:
The ontology of bohmian mechanics.
The British Journal for the Philosophy of Science
(2014)
\end{botherref}
\endbibitem

\bibitem[\protect\citeauthoryear{Deckert et~al.}{}]{deckert22}
\begin{botherref}
\oauthor{\bsnm{Deckert}, \binits{D.}},
\oauthor{\bsnm{D{\"u}rr}, \binits{D.}},
\oauthor{\bsnm{Pickl}, \binits{P.}}:
22 bohmian grids and the numerics of schr{\"o}dinger evolutions
\end{botherref}
\endbibitem

\bibitem[\protect\citeauthoryear{Oldofredi}{2022}]{Oldofredi2022}
\begin{bbook}
\bauthor{\bsnm{Oldofredi}, \binits{A.}}:
In: \beditor{\bsnm{Allori}, \binits{V.}} (ed.)
\bbtitle{Beables, Primitive Ontology and Beyond: How Theories Meet the World},
pp. \bfpage{97}--\blpage{111}.
\bpublisher{Springer},
\blocation{Cham}
(\byear{2022}).
\doiurl{10.1007/978-3-030-99642-0_7} .
\burl{\url{https://doi.org/10.1007/978-3-030-99642-0_7}}
\end{bbook}
\endbibitem

\bibitem[\protect\citeauthoryear{Maudlin}{2013}]{maudlin2013nature}
\begin{botherref}
\oauthor{\bsnm{Maudlin}, \binits{T.}}:
The nature of the quantum state.
The wave function: Essays on the metaphysics of quantum mechanics,
126--53
(2013)
\end{botherref}
\endbibitem

\bibitem[\protect\citeauthoryear{Maudlin}{2019}]{maudlin2019philosophy}
\begin{bbook}
\bauthor{\bsnm{Maudlin}, \binits{T.}}:
\bbtitle{Philosophy of Physics: Quantum Theory}.
\bpublisher{Princeton University Press},
\blocation{Princeton}
(\byear{2019})
\end{bbook}
\endbibitem

\bibitem[\protect\citeauthoryear{De~Broglie}{1958}]{de1958tentative}
\begin{botherref}
\oauthor{\bsnm{De~Broglie}, \binits{L.}}:
Une tentative d'interpr{\'e}tation causale et non lin{\'e}aire de la m{\'e}canique ondulatoire.
British Journal for the Philosophy of Science
\textbf{9}(34)
(1958)
\end{botherref}
\endbibitem

\end{thebibliography}

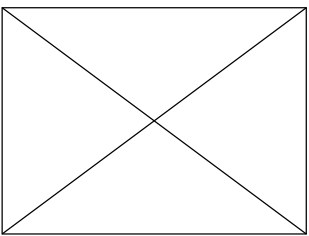
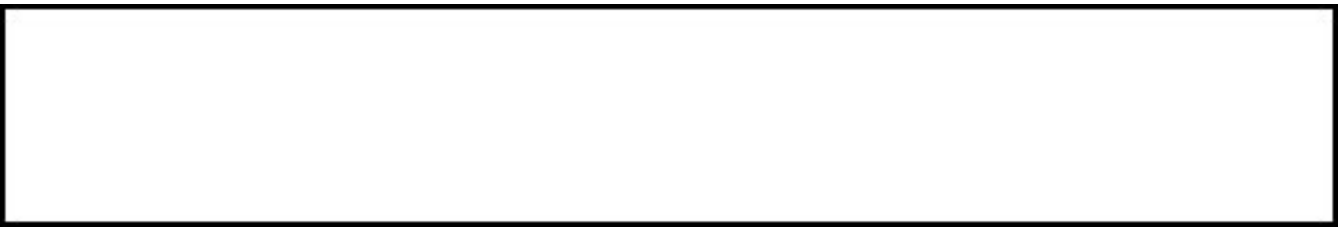
\end{document}